\documentclass[final,5p,times,twocolumn,sort&compress]{elsarticle}
\usepackage{amsmath,latexsym,mathtools}
\usepackage[x11names,dvipsnames,table]{xcolor} 

\usepackage{enumitem}

\usepackage{subcaption}
\usepackage{changepage}
\usepackage{xcolor}
\usepackage{float}

\usepackage[colorlinks]{hyperref}
\usepackage[thinc]{esdiff}

\AtBeginDocument{\hypersetup{
	colorlinks=true,
	linkcolor=black!40!blue,
	citecolor=black!40!blue,
	filecolor=black,
	urlcolor=black!45!blue,
}}

\newcommand{\X}{\mathcal{X}} 
\newcommand{\K}{\mathcal{K}}

\newcommand{\change}[1]{\textcolor{black}{#1}}

\newcommand{\orcidicon}{\includegraphics[width=0.32cm]{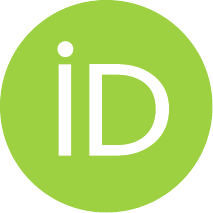}}

\newcommand\orcidT{{\href{https://orcid.org/0000-0001-8007-5181}{\orcidicon}}}
\newcommand\orcidM{{\href{https://orcid.org/0000-0002-1529-1889}{\orcidicon}}}
\newcommand\orcidP{{\href{https://orcid.org/0000-0001-8073-4896}{\orcidicon}}}

    \usepackage[pscoord]{eso-pic}
   \newcommand{\placetextbox}[3]{
	\setbox0=\hbox{#3}
	\AddToShipoutPictureFG*{
		\put(\LenToUnit{#1\paperwidth},\LenToUnit{#2\paperheight}){\vtop{{\null}\makebox[0pt][c]{#3}}}%
	}%
}%

\begin{document}

    \title{Dark energy with a shift-symmetric scalar field: obstacles, loophole hunting and dead ends}

\author[1]{Teodor Borislavov Vasilev\orcidT\corref{cor1}}
\ead{teodorbo@ucm.es}

\author[2,3]{Mariam Bouhmadi-López\orcidM}
\ead{mariam.bouhmadi@ehu.eus}

\author[1]{Prado Martín-Moruno\orcidP}
\ead{pradomm@ucm.es}

\cortext[cor1]{Corresponding author}

    \affiliation[1]{organization={Departamento de Física Teórica and IPARCOS, Universidad Complutense de Madrid},
    addressline={Pl. de las Ciencias 1},
    postcode={ES 28040},
    city={Madrid},
    country={Spain.}}

    \affiliation[2]{organization={IKERBASQUE, Basque Foundation for Science},
    postcode={ES 48011},
    city={Bilbao},
    country={Spain.}}

    \affiliation[3]{organization={Department of Physics and EHU Quantum Center, University of the Basque Country, UPV/EHU},
    addressline={P.O. Box 644},
    postcode={ES 48080},
    city={Bilbao},
    country={Spain.}}

    \journal{Physics of the Dark Universe}

 
\begin{abstract}
    We discuss the possibility of a scalar field being the fundamental description of dark energy. We focus on shift-symmetric scalar-tensor theories \change{since this symmetry potentially avoids some fine-tuning problems. We also restrict attention to theories} satisfying that the propagation speed of gravitational waves is equal to the speed of light. \change{These considerations lead us to investigate shift-symmetric Kinetic Gravity Braiding theories.}
    Analysing the stability of scalar linear perturbations, we discuss the conditions that seems to be necessary to describe (super) accelerated cosmic expansion without introducing instabilities. \change{However,
    it has been previously established that the linearised analysis does not guarantee the stability of this non-canonical scalar theory, as potentially dangerous interactions between dark energy fluctuations and tensor perturbations (essentially gravitational waves) appear at a higher order in perturbation theory.} 
    Indeed, although we shall point out that the standard proof of absence of dark energy stable \textit{braiding} models due to this interaction has a possible way-out, we find general arguments suggesting that there are no dark energy stable solutions that can exploit this loophole. Thus, we discuss future research directions for finding viable fundamental descriptions of dark energy. We also provide a dictionary between the covariant version of the scalar field theory and  that of the Effective Field Theory approach, explicitly computing the parameters in the latter formalism in terms of the functions appearing in the covariant version, and its derivatives.
    To the best of our knowledge, this is the first time these expressions are explicitly obtained up-to arbitrary order in perturbation theory.
\end{abstract}

\maketitle

    \placetextbox{0.85}{0.95}{IPARCOS-UCM-24-030}

\section{Introduction}

Scalar field theories could be crucial in providing an underlying theoretical framework for addressing the mysteries of dark energy (DE). These theories introduce a scalar field that can be used to describe dynamical DE, in contrast to the cosmological constant case (see, for instance, references \cite{Peebles:2002gy,Copeland:2006wr,Bamba:2012cp}). To select the appropriate scalar field theory for cosmology, it is reasonable to demand that the corresponding field equations to be of second order. Otherwise, the new degrees of freedom (DOF) may destabilise the dynamics of the theory by means of the Ostrogradski ghost \cite{Woodard:2015zca}. In this context, a natural framework is that provided by Horndeski theory \cite{Horndeski:1974wa} (see also \cite{Kobayashi:2019hrl} for a review), although some theories with higher-order derivatives in the action can also avoid the Ostrogradski instability through degeneracy conditions in the Lagrangian (see, for instance, \cite{Langlois:2015cwa,BenAchour:2016cay}). 
Rather recently, the speed of propagation of gravitational waves (GWs) has also become an important consideration when formulating theories beyond general relativity. After the detection of the event GW170817 \cite{LIGOScientific:2017vwq} and its electromagnetic counterpart \cite{Goldstein:2017mmi}, it has been concluded that the propagation speed of GWs is very close to that of light.
\change{Selecting the Horndeski models that trivially propagate GWs at the speed of light reduces the original parameter-space of the theory to the so-called viable subclass of Horndeski \cite{Kobayashi:2019hrl}. This includes the well-known $k$-essence scenario \cite{Armendariz-Picon:1999hyi,Armendariz-Picon:2000nqq,Chiba:1999ka,Scherrer:2004au,dePutter:2007ny}, the more general Kinetic Gravity Braiding (KGB) theory \cite{KGB1}, and a possible non-minimal coupling to gravity \textit{via} the scalar field (see, for instances,  reference \cite{Kase:2018aps}).}

Scalar fields minimally coupled to gravity and with a canonical kinetic term have been already used to describe the current accelerated expansion of the Universe. However, in that simple case one has to rely on the particular form of a potential term, re-introducing the fine-tuning problem usually associated with the cosmological constant. This issue can be avoided by focusing on models that are invariant under constant shifts in the scalar field, for which potential-like terms do not appear in the Lagrangian. This symmetry-based argument is an elegant way to make the evolution of the system to depend only on the rate of change of the scalar field but not on the scalar field itself (see discussion in reference \cite{KGB1}). 
However, the simplest minimally coupled scalar field with a canonical kinetic term in the action is only able to describe a stiff fluid when imposing shift-symmetry (see, for example, references \cite{Copeland:2006wr,Tsujikawa:2013fta}); leading to a non-interesting phenomenology for the (late-time) DE sector. 
On the other hand, in the Horndeski subfamily previously mentioned this argument prevents from having a non-minimal coupling to gravity \textit{via} the scalar field itself, although a derivative coupling to matter is still allowed in principle. In either case, evading fifth-force constraints coming from astrophysical scales is naturally easy under this symmetry consideration \cite{KGB1}. The Horndeski models that satisfy this symmetry condition and predict a propagation speed for GWs equal to the speed of light are known as shift-symmetric KGB theories \cite{KGB1}.

Shift-symmetric KGB theories are well-known for producing self-tuning de-Sitter future solutions \cite{KGB1} (see also references \cite{DeFelice:2010pv,Bernardo:2021hrz,Germani:2017pwt,Martin-Moruno:2015bda,Martin-Moruno:2015kaa,DeFelice:2011bh,Tsujikawa:2010zza}), but they can also describe different future phenomenology  \cite{Vasilev:2022wiv,BorislavovVasilev:2022gpp}. \change{Contrary to the $k$-essence scenario, in the KGB theory second order derivatives of the metric are present in the scalar field equation of motion and vice versa, second order derivatives of the scalar field appear in the metric field equation \cite{KGB1}. As a result, these theories can describe a phantom regime that is stable in the sense of no ghost- or gradient-like instabilities for linear scalar perturbation \cite{KGB1}. 
In fact, the KGB theory can also produce a smooth phantom crossing in the DE sector \cite{KGB1}, a feature not possible with a single scalar field \textit{à la} $k$-essence \cite{Vikman:2004dc,Caldwell:2005ai}. With regard to phantom energy,} it should be highlighted that the violation of the Null Energy Condition (NEC) in the DE sector is not only still allowed by the observational data (see discussion in reference \cite{Risaliti:2018reu}), but it has been indeed proven that a phantom regime (with phantom crossing) is a necessary prerequisite to ease both the $H_0$ and $S_8$ cosmological tensions simultaneously by taking into account new physics that is relevant only at late cosmic times \cite{Heisenberg:2022lob,Heisenberg:2022gqk}. 
Therefore, the motivation, simplicity and apparent stability properties of the shift-symmetric KGB scalar field models makes this framework an interesting proposal for modelling DE. 
Nevertheless, the absence of instabilities in linear scalar perturbations is a necessary but not a sufficient condition for the stability of the classical theory. Indeed, it has been shown that the interaction mediated by the braiding term between tensor perturbations (essentially GWs) and DE fluctuations may induce a ghost-like and/or gradient-like instability in the scalar sector \cite{Creminelli3}. (Se also references \cite{Creminelli1,Creminelli2} for a discussion on the decay of GWs into scalar fluctuations when Lorentz invariance is spontaneously broken.) Consequently, it was concluded that the braiding term in the KGB theory should not produce any sizeable effect in order to trivially escape from the GWs-induced instabilities \cite{Creminelli3}. According to these results, the available parameter-space of the shift-symmetric KGB models seems to shrink to that of the shift-symmetric $k$-essence theory only \cite{Creminelli3}. 

In the present work, we first re-analyse the inter-relation between phantom behaviour and classical and semi-classical instabilities of the scalar field in the framework of shift-symmetric KGB models. By implementing a novel characterization of the already-known stability conditions, we shed some light on previous results in the literature for $k$-essence models  \cite{Creminelli:2008wc} (see also, for instance, references \cite{Armendariz-Picon:1999hyi,Armendariz-Picon:2000nqq,Garriga:1999vw,Armendariz-Picon:2000ulo,Hsu:2004vr}) and also obtain new compact results for the braiding term. In the second place, we reflect about the instability induced by the interaction of the braiding term with GWs \cite{Creminelli3}.
We propose a possible way to circumvent this obstacle when departing from the Galileon terms for the KGB functions. Unfortunately, we shall argue that the apparent loophole seems to lack practical applicability for constructing viable dark energy models with a non-vanishing braiding term. In order to perform this study, we have derived the detailed form of the mass parameters of the KGB theory in the Effective Field Theory (EFT) approach. Although the expressions for these parameters were already computed at leading order, to the best of our knowledge, this is the first time that the complete family of parameters up-to arbitrary order is presented.

This work is organised as follows: we review the general properties of shift-symmetric KGB theories in the first part of section \ref{sec:linearPerturbs}. The stability of linear scalar perturbations in the well-known $k$-essence subclass of the more general KGB framework is reviewed in section \ref{sec:kessence}. In sections \ref{sec:PB} and \ref{sec:fullKGB} we discuss the viability of linearised scalar perturbation for the remaining part of the KGB theory. Section \ref{sec:intGWs} is devoted to the interaction of the braiding term with GWs. We review the arguments presented in reference \cite{Creminelli3} in section \ref{sec:revGWs}. We identify a potential mechanism to ease the constraints found in reference \cite{Creminelli3} in section \ref{sec:noGrad}. In section \ref{sec:noExample} we discuss that the possibility for a viable KGB model to implement such a mechanism is an elusive ambition in practice. Therefore, in section \ref{sec:conclusions}, we reflect about the open paths that can still be further explored to find fundamental viable descriptions for dark energy. Finally, auxiliary results for discussing the stability of linear perturbations are summarised in  \ref{app:FKGB}.  A self-contained guide to the EFT approach to DE and useful calculations are presented in \ref{app:EFT}.


\section{Stability of linearised scalar perturbations\label{sec:linearPerturbs}} 

    The KGB theory is given by the action \cite{KGB1} (see also \cite{Kobayashi:2010cm,Bellini:2014fua})
	\begin{eqnarray}\label{eq:actionKGB}
		S_g=\int d^4 x\sqrt{-g}\left[\frac12 R+ K(\phi,X)-G(\phi,X)\Box\phi \right],
	\end{eqnarray}
	where we have adopted the geometric unit system $8\pi G=c=1$, $K(\phi,X)$ and $G(\phi,X)$ are arbitrary functions of the scalar field $\phi$ and its canonical kinetic term $X\coloneqq-\frac12 g^{\mu\nu}\nabla_\mu\phi\nabla_\nu\phi$,
    the box represents the covariant d'Alembertian operator $\Box \phi=g^{\mu\nu}\nabla_\mu\nabla_\nu\phi$ and we have used the signature $(-,+,+,+)$. 

    \change{It should be noted that one of the arguments most commonly used to go beyond the standard model of cosmology is the existence of a fine-tuning problem regarding the value of the cosmological constant. However, having potential terms in a scalar-tensor theory could reintroduce the fine-tuning problem now regarding the parameters that fix the minimum of the potential, if the accelerated expansion is reached in the potential domination regime. One possible way of trying to avoid this problem is the consideration of a shift-symmetric scalar field, which guarantee  that the cosmic acceleration is driven by the kinetic energy. Those are models} that are invariant under constant shifts in the scalar field, for which potential-like terms do not appear in the Lagrangian. This symmetry-based argument is an elegant way to make the evolution of the system to depend only on the rate of change of the scalar field but not on the scalar field itself (see discussion in reference \cite{KGB1}). 
    Therefore, in this work we will focus on the shift-symmetric sector of the KGB theory \cite{KGB1} (see also \cite{DeFelice:2010pv,Bernardo:2021hrz,Germani:2017pwt,Martin-Moruno:2015bda,Martin-Moruno:2015kaa,DeFelice:2011bh,Tsujikawa:2010zza,Vasilev:2022wiv,BorislavovVasilev:2022gpp}). This is when the above action is invariant under constant shifts in the scalar field given by
    \begin{align}
        \phi\to\phi+c,
    \end{align}
    being $c$ some constant. In practice, this implies that the functions $K$ and $G$ do not depend on the scalar field but only on its kinetic term. Hence, we will focus on the particular case of $K=K(X)$ and $G=G(X)$. (Nevertheless, this assumption will be relaxed when discussing the EFT approach in  \ref{app:EFT}.)
    In addition, we will also consider the spatially flat cosmological background described by the homogeneous and isotropic Friedmann-Lema\^{i}tre-Robertson-Walker (FLRW) space-time. That corresponds to the line element
  	\begin{align}
  		ds^2=-dt^2+a^2(t)\,dx_3^2,
  	\end{align}
    where  $a(t)$ stands for the scale factor and $dx_3^2$ are the spatial three-dimensional flat sections. Taking matter and radiation as external sources to the action (\ref{eq:actionKGB}), the field equations of the theory for this background read \cite{KGB1}
	\begin{align}
		3H^2=&\,\rho_m+\rho_r-K+\dot{\phi}J,\label{FE1}\\
		\dot{H}=&-\frac12\left(\rho_m+\frac43\rho_r\right)+XG_X \ddot{\phi}-\frac12\dot{\phi}J,\label{FE2}\\
		\dot{\rho}_m=&-3H\rho_m,\label{eq:conservationDust}\\
		\dot{\rho}_r=&-4H\rho_r\label{eq:conservationRad},
	\end{align}
    where a dot represents derivation w.r.t. the cosmic time $t$ and $G_X$ stands for ${\rm d}G/{\rm d}X$. In addition,
    \begin{align}\label{eq:J}
		J(\dot\phi,H)\coloneqq \dot{\phi}K_X+6HXG_X,
    \end{align}
    is the shift-current related to the global shift-symmetry of the theory \cite{KGB1}.
    In view of these equations, the energy density and pressure of the dark fluid can be defined as
    \begin{equation}\label{rho_phi}
        \rho_\phi(\dot\phi,H)\coloneqq \dot\phi J-K,
    \end{equation}
    and
    \begin{equation}\label{p_phi}
        p_\phi(\dot\phi,\ddot\phi)\coloneqq K-2XG_X\ddot\phi,
    \end{equation}
    respectively. Then, the equation of state parameter and the partial energy density for the scalar field take the form $w_\phi\coloneqq p_\phi/\rho_\phi$ and $\Omega_\phi\coloneqq\rho_\phi/(3H^2)$, respectively.
    \change{Contrary to $k$-essence \cite{Armendariz-Picon:1999hyi},} the energy density, $\rho_\phi$, now depends on the Hubble parameter through $J$ while the pressure, $p_\phi$, contains $\ddot\phi$ \cite{KGB1}. Moreover, the pressure will also show an explicit dependence on the Hubble function \change{(and the external sources)} when  $\ddot\phi$ is removed using the field equations.
    
    The field equation of the scalar field can be obtained varying the action (\ref{eq:actionKGB}) w.r.t. the scalar field. This yields
     \begin{align}
          A(\dot{\phi},H)\ddot{\phi}+6XG_X \dot{H}+3HJ=0, \label{SFIELD}
    \end{align}
     where
    \begin{align}
		A(\dot{\phi},H)\coloneqq K_X+2XK_{XX}+6H\dot{\phi} \left(G_X+XG_{XX}\right),\label{eq:A}
    \end{align}
    has been defined for the compactness of the notation. This equation is completely equivalent to the covariant conservation of the $\phi$-fluid in the hydrodynamic approach, \change{ that is
    \begin{align}\label{SFIELD_hydro}
        \dot\rho_\phi+3H(\rho_\phi+p_\phi)=0,
    \end{align}
    where $\rho_\phi$ and $p_\phi$ correspond to the scalar field energy density and pressure defined in expressions (\ref{rho_phi}) and (\ref{p_phi}), respectively. It should be noted, however, that the scalar field equation of motion (either in the form of equation (\ref{SFIELD}) or, equivalently, in the hydrodynamic approach depicted in equation (\ref{SFIELD_hydro}))} is not independent from the system of equations given in expressions (\ref{FE1}) to (\ref{eq:conservationRad}).

    Alternatively to the scalar field equation (\ref{SFIELD}), one can obtain the same information from the conservation equation related to the shift-symmetry of the theory; that is
    \begin{align}
        \frac{1}{a^3}\diff{\left(a^3J\right)}{t}=0.\label{eq:phi}
    \end{align}
    As noted in the reference \cite{KGB1}, this conservation trivially leads to a first integral of motion for the system  given by
    \begin{align}
        J=\frac{Q_0}{a^3},
    \end{align}
    where $Q_0$ is a constant. This feature of the shift-symmetric KGB theories can be used to significantly simplify the discussion on the evolution of these cosmological models \cite{KGB1}. This is because $J=0$ defines a surface in the configuration space towards which all trajectories evolve if the scale factor is growing large. Therefore, the locus $J=0$ can be used to simplify the field equations, at least asymptotically. 
    \change{ However, this simplification should be taken with due caution: the vanishing of the current $J$ may not always imply that the term $\dot\phi J$ should be dropped from equations (\ref{FE1}) and (\ref{FE2}); see discussion in references \cite{Vasilev:2022wiv,BorislavovVasilev:2022gpp}.}
    In general, the future phenomenology of this shift-symmetric scalar field theory has been shown to encompass a great variety of options. From self-tuning de-Sitter-like future solutions \cite{KGB1} (see also references \cite{DeFelice:2010pv,Bernardo:2021hrz,Germani:2017pwt,Martin-Moruno:2015bda,Martin-Moruno:2015kaa,DeFelice:2011bh,Tsujikawa:2010zza}) to DE-driven cosmological singularities \cite{Vasilev:2022wiv,BorislavovVasilev:2022gpp}.

    The stability of scalar perturbations around a FLRW background have already been addressed at linear order for this scalar field theories \cite{KGB1} (see also references \cite{DeFelice:2011bh,Bellini:2014fua,Pujolas:2011he}). 
    \change{Indeed, the quadratic action describing scalar perturbations reads \cite{KGB1,DeFelice:2011bh}
     \begin{align}\label{eq: action LPerturbations}
        &S^{\rm (2)}_{\rm scalar} =\int \textup{d}^4x \, a^3 Q_S\left[\dot{\zeta}^2-\frac{c_s^2}{a^2}\left({\partial_i} \zeta\right)^2\right],
    \end{align}
    where $\zeta$ represents the curvature perturbation as commonly defined (see, for example, reference \cite{KGB1}), $Q_{\rm S}$ is the normalization factor of the kinetic term, and $c_s$ denotes the speed of propagation (hereon referred to as the ``speed of sound'') of the scalar perturbation.}
    From the above action, the absence of ghost and gradient instabilities in the scalar sector implies \cite{KGB1,Bellini:2014fua}
    \begin{align}
		Q_{\text{S}} & \coloneqq\frac{2D}{(2-\alpha_{\textrm{B}})^{2}}>0\,,\qquad 
        D\coloneqq\alpha_{\textrm{K}}+\frac{3}{2}\alpha_{\textrm{B}}^{2}\,,\label{eq:scalarstab}\\
		c_{\text{s}}^{2} &\coloneqq-\frac{\left(2-\alpha_{\textrm{B}}\right)\left[\dot{H}-\frac{1}{2}H^{2}\alpha_{\textrm{B}}\right]-H\dot{\alpha}_{\textrm{B}}+\rho_{\textrm{m}}+\frac43\rho_{\textrm{r}}}{H^{2}D}\geq0,\label{eq:gradstab} 
    \end{align}
    respectively, where the dimensionless functions $\alpha_K$ and $\alpha_{\textrm{B}}$ were first introduced in reference \cite{Bellini:2014fua}. They are defined as
    \begin{align}\label{eq:alphaK}
		H^{2}\alpha_{\textrm{K}}\coloneqq &\, 2X\left(K_{X}+2XK_{XX}\right) +12\dot{\phi}XH\left(G_{X}+XG_{XX}\right), \\
		H\alpha_{\textrm{B}}\coloneqq &\, 2\dot{\phi}XG_{X}.\label{eq:alphaB}
    \end{align}
    \change{Note that condition (\ref{eq:scalarstab}) trivially reduces to $D>0$, since the denominator of $Q_{\rm S}$ is always positive.
   Violation of this condition leads} to scalar perturbations whose associated kinetic energy takes negative values  \cite{Rubakov:2014jja}; i.e. the perturbations become ghost-like. The presence of a ghost DOF may be detrimental (at the classical level) if the ghost interacts with a positive-energy mode, as this may lead to runaway solutions where the total energy is conserved but the relative energies associated to the ghost and no-ghost sectors diverge (see, for instance, reference \cite{Carroll:2003st}). At the quantum level, the corresponding $\phi$-particles would have negative energies. 
    In this case the vacuum becomes unstable due to the spontaneous production of ghost-particles together with normal-particles with arbitrarily high energies and momenta \cite{Carroll:2003st} (see also reference \cite{Rubakov:2014jja}). Consequently, backgrounds with ghosts are generally considered  pathological at both classical and quantum levels. 
    On the other hand, violation of condition (\ref{eq:gradstab}) introduces the so-called gradient instability. This is when the leading order spatial derivatives have the wrong sign w.r.t. the time derivatives in the perturbed action (\ref{eq: action LPerturbations})  (see also references \cite{Rubakov:2014jja,Frusciante:2019xia}). In Fourier space, the frequencies of the oscillations become imaginary at high momenta, resulting in perturbations that grow exponentially fast. Thus, it precludes a stable classical model (at least, as long as a perturbative treatment is still valid).

    It is interesting to note that
	\begin{align}\label{eq:ghost}
		H^2D=2X\left(A+6X^2G_X^2\right),
	\end{align}
    and, therefore, the ghost-free condition $D>0$ (see equation (\ref{eq:scalarstab})) implies
    \begin{align}
		A+6X^2G_X^2>0,
    \end{align}
    in our notation. Moreover, replacing the time derivative of the Hubble rate that appears in the scalar field equation (\ref{SFIELD}) by means of the Raychaudhuri equation (\ref{FE2}) leads to 
    \begin{align}\label{SFIELDreduced}
        H^2D\ddot{\phi}+6XJ\left(H-\dot{\phi}XG_X\right)-2X^2G_X\left(4\rho_r+3\rho_m\right)=0.
    \end{align}
    Hence, the ghost parameter $D$ is just the factor ahead of the scalar field acceleration.
    
    In the following we will analyse the stability conditions described in equations (\ref{eq:scalarstab}) and (\ref{eq:gradstab}) for three different scenarios:  $(i)$ the $k$-essence model ($G\equiv0$), $(ii)$ when $K\equiv 0$, and $(iii)$ the most general KGB scenario. These are sections \ref{sec:kessence}, \ref{sec:PB} and \ref{sec:fullKGB}, respectively. In doing so we will review some well-known facts and also face new discussions about the viability of these shift-symmetric theories from the point of view of linear cosmological perturbations. 


\subsection{K-essence\label{sec:kessence}}

    Let us first re-analyse the stability for the well-known $k$-essence case \cite{Armendariz-Picon:1999hyi,Armendariz-Picon:2000nqq,Chiba:1999ka,Scherrer:2004au,dePutter:2007ny}. 
    \change{This corresponds to setting the braiding function $G$ to zero in action (\ref{eq:actionKGB}), although a non-zero but constant braiding function would also reduce to the $k$-essence scenario. From this point of view, one may  consider the $k$-essence theory as a special sub-case of the more general KGB action (\ref{eq:actionKGB}). However, it is worth noting that the two scalar field theories have very different properties. This is because second order derivatives of the metric and scalar field are no longer mixed when $G$ is constant. 
    Thus, some of the most interesting features of the KGB set-up (such as stable phantom crossing) are not present in the $k$-essence scenario. Still, $k$-essence represents the most general scalar theory whose action contains up-to first order derivatives of the scalar field and has been extensively explored in a wide variety of cosmological scenarios, from inflation \cite{Armendariz-Picon:1999hyi,Garriga:1999vw} to DE \cite{Armendariz-Picon:2000nqq,Chiba:1999ka,Scherrer:2004au,dePutter:2007ny}.
    }
    
    Unlike the scalar field with a canonical kinetic term, for a $k$-essence field a phantom behaviour is not necessarily due to the presence of a ghost \cite{Hsu:2004vr} (see also \cite{Creminelli:2008wc}), since equation (\ref{eq:scalarstab}) reduces to
    \begin{equation}\label{KQ}
       Q_{\text{S}} =\frac{D}{2}=\frac{X}{H^2}\diff{\rho_\phi}{X},
    \end{equation}
    whereas the violation of the NEC by the scalar field depends on the slope of the $k$-essence function $K$,
     \begin{equation}\label{Kw}
        1+w_\phi=\frac{2XK_X}{\rho_\phi},
    \end{equation}
    being
    \begin{align}
        \rho_\phi=&\,2XK_X-K,\\
        p_\phi=&\,K,
    \end{align}
    cf. with expressions (\ref{rho_phi}) and (\ref{p_phi}) when $G_X$ is trivial. The slope of the function $K$ should be negative for a phantom field, but this do not necessarily force the energy density $\rho_\phi$ to be a decreasing function of the kinetic term. So, the sign on the r.h.s. of equation (\ref{KQ}) is in principle independent of that of equation (\ref{Kw}); see reference \cite{Hsu:2004vr}. However, the relation between instabilities and phantom behaviour appears when considering the squared speed of sound of the perturbations, i.e. $c_s^2$. Taking into account the Friedmann equation (\ref{FE2}), we can recast the gradient-free condition (\ref{eq:gradstab}) in term of the background quantities as
	\begin{align}\label{eq:gradstabK}
		c_s^2=\frac{3(1+w_\phi)\,\Omega_\phi}{D}\geq0,
	\end{align}
    where $\Omega_\phi\coloneqq\rho_\phi/3H^2$ is the partial energy density for the scalar field and we have also used the fact that $\alpha_{\textrm{B}}\equiv0$ for $k$-essence.
    This new reformulation of the expression for $c_s^2$ is extremely useful since we can address the stability of linear perturbations in terms of the background quantities for which we have a strong physical intuition.  Since we are focusing on the case $\Omega_\phi>0$, this inequality shows an inter-relation between the ghost- and gradient-free conditions with the possible phantom character of the $\phi$-fluid. In particular, if the NEC is satisfied for the scalar field (i.e. if $1+w_\phi>0$), there is not a gradient instability if and only if the ghost-free condition is also satisfied. As a result, there are $k$-essence models with a completely stable scalar spectrum. On the other hand, the gradient-free and ghost-free conditions are anti-correlated when the NEC is violated for the $\phi$-fluid. That is to say, $c_s^2$ is positive for a phantom fluid only if a ghost-like instability is present and vice-versa, scalar perturbations are ghost-free only if a gradient instability is triggered. Please note that the same conclusion was reached in reference \cite{Creminelli:2008wc}, albeit through a different reasoning (see also references \cite{Hsu:2004vr,Abramo:2005be}).

    As is well-known, it is not possible to describe an effective phantom fluid via a scalar field \textit{à la} $k$-essence in a way that both ghost and gradient instabilities are absent \cite{Hsu:2004vr,Abramo:2005be,Creminelli:2008wc,Rubakov:2014jja}. However, let us reflect about the nature of these instabilities (see, for example, reference \cite{Rubakov:2014jja} for a review on the topic). On the one hand, the presence of gradient instabilities is related to a wrong sign for the spatial derivatives in the perturbed action (\ref{eq: action LPerturbations}). This makes the frequencies of the fluctuations imaginary at high momenta, resulting in perturbations that grow arbitrary fast. Thus, it precludes a stable classical model (at least, as long as a perturbative treatment is still valid). 
    For a ghost mode, on the other hand, the associated kinetic energy is negative. If the ghost DOF interacts with a positive energy mode (at the classical level) there may be runaway solutions, where the total energy is conserved while individual energies diverge; we referred the interested reader to the discussion in, for example, reference \cite{Carroll:2003st}. Thereby, the common lore states that ghost instabilities are catastrophic (already) at classical level. Still, the classical background is stable against high momenta perturbations \cite{Rubakov:2014jja}.  Nevertheless, it is interesting to note that there are examples in the literature in which the presence of a ghost DOF interacting with a positive energy DOF do not lead to runaway solutions and, therefore, to the destabilization of the classical motion of the system \cite{Deffayet:2021nnt}. (See also ``islands of stability'' \cite{Pagani:1987ue,Smilga:2004cy} and meta-stability of ghosts \cite{Salvio:2019ewf,Gross:2020tph}.)
    Upon canonical quantisation of theories with ghosts, the energy conservation does not forbid pair creation from vacuum of ghosts-particles together with normal-particles: the vacuum becomes quantum mechanically unstable (see, for instance, references \cite{Rubakov:2014jja,Carroll:2003st}). So, the presence of a ghost is also considered as pathological at the semi-classical level. However, the possibility of safely living with ghosts at quantum level has already been explored with positive conclusions \cite{Hawking:2001yt,Garriga:2012pk,Salvio:2015gsi,Smilga:2017arl} \change{(see also references \cite{Salvio:2020axm,Salvio:2024joi} for applications to quadratic gravity)}. 
    Therefore, even if it is clear that the presence of any of these instabilities (ghost and/or gradient) is a red light that has to be seriously taken into account, it would be interesting to consider the possibility that the scalar field (understood as being of gravitational nature) should be quantized in a different way to matter fields (for example, with the space-time itself \cite{Dabrowski:2006dd,Bouhmadi-Lopez:2009ggt}) or not been quantized at all, when dealing with the ghost instability. For a review on the topic see, for example, reference \cite{Bouhmadi-Lopez:2019zvz}.
    
    In this context, a gradient instability is hardly tameable in the classical theory, whereas a ghost-like instability could be further discussed at both classical and quantum levels.  So, if one considers that the $H_0$ tension has to be solved by means of a phantom fluid, then, our best hope in the $k$-essence case would be selecting the function $K$ in such a way that only the ghost-free condition is violated. Otherwise, a negative $c_s^2$ will inevitably jeopardise the perturbative approximation due to fast growing solutions.
    On the other hand, it is also worth to mention that the scalar field in (shift-symmetric or not) $k$-essence theory cannot cross the phantom divide \cite{Vikman:2004dc} (see also \cite{Caldwell:2005ai}). The equation of state parameter for the scalar field, namely $w_\phi$, is always greater or less than $-1$ for a single scalar field \textit{à la} $k$-essence. Nevertheless, a phantom-crossing in the DE sector was analytically shown to be a prerequisite for solving both the $H_0$ and $S_8$ tensions simultaneously \cite{Heisenberg:2022lob,Heisenberg:2022gqk}. (Similar conclusions were also reached by addressing the resolution of the $H_0$ tension only \cite{Banerjee:2020xcn,Lee:2022cyh}.) Therefore, if both tensions were considered on the same footing\footnote{Note that the disagreement between the value of $H_0$ inferred from Planck-Cosmic Microwave Background \cite{Planck:2018vyg} and that coming from the direct local distance ladder measurements provided by the SH0ES team \cite{Riess:2021jrx,Riess:2022mme} has reached a statistical significance of more than 5$\sigma$. Whereas the $S_8$ tension is at much lower statistical significance, reaching between 1.7 to 3 standard deviations; see for instance  \cite{Abdalla:2022yfr} and references therein. We refer the interested reader to the review \cite{Perivolaropoulos:2021jda} for a self-contained discussion about the observational tensions in the standard cosmological model.}, a scalar field theory beyond $k$-essence should be explored. See also references \cite{Banerjee:2020xcn,Lee:2022cyh} for previous works on the $H_0$ tension.

\subsection{Braiding model\label{sec:PB}}

    \change{ We now analyse the case of $K\equiv0$ in action (\ref{eq:actionKGB}). This is a straightforward limiting scenario of the complete KGB framework where the contribution of the $k$-essence part is negligible. However, contrary to pure $k$-essence discussed in the previous section, second order derivatives of the metric and scalar field are still mixed in the corresponding field equations due to the function $G$; see equations (\ref{FE2}) and (\ref{SFIELD}). 
    As a result, this is a scenario simpler than the complete KGB theory that still provides us with an interesting candidate for addressing the stability properties of the \textit{kinetic braiding}.
    }
    
    The expressions for the energy density and pressure of the scalar field in equations (\ref{rho_phi}) and (\ref{p_phi}), respectively, simplify to
    \begin{align}
        \rho_\phi&=6\dot{\phi}XG_XH,\\
        p_\phi&=-2XG_X\ddot{\phi},
    \end{align}
    where the expression for $J$ in equation (\ref{eq:J}) has been also considered. Then, the equation of state parameter for the scalar field reads
    \begin{align}
        w_\phi=-\frac{\ddot{\phi}}{3\dot{\phi}H}.
    \end{align}
    The scalar field acceleration, $\ddot{\phi}$, can be removed from the previous expression by means of the Raychaudhuri and scalar field equations, these are expressions (\ref{FE2}) and (\ref{SFIELD}), respectively. Then, after simple manipulations we can re-write the scalar field equation of state as
    \begin{align}
		w_\phi&=\frac{\Omega_\phi(3-\Omega_r)}{2D},
    \end{align}
    where $\Omega_r\coloneqq\rho_r/3H^2$ is the partial energy density for radiation and the contribution of matter has been removed through the constraint $\Omega_m=1-\Omega_r-\Omega_\phi$ coming form the Friedmann equation (\ref{FE1}).
    Please note that the above expression allow us to re-express the ghost-free condition (\ref{eq:scalarstab}) in terms of background quantities as
    \begin{align}\label{eq:ghostPB}
		D=\frac{\Omega_\phi(3-\Omega_r)}{2w_\phi}>0.
    \end{align}
    As emphasised in the previous section, this new approach to the evaluation of the stability conditions is extremely useful since we have a strong physical intuition for the values of  background quantities.
    The functional form of the equation of state for the scalar field, that is to say $w_\phi=w_\phi(a,\dot{a},\dot\phi,\ddot\phi)$, will depend on the choice for the braiding function $G$. However, if we assume a positive $\Omega_\phi$, it follows from the above inequality that a ghost instability is always present if $w_\phi$ is negative. 
    A significant implication of this result is that any model that aims to describe a DE component (not necessary phantom-like) will inevitably have this instability, since   $w_\phi$ should be less than $-1/3$.

    \begin{figure}[t]
		\centering
		\includegraphics[scale=0.7]{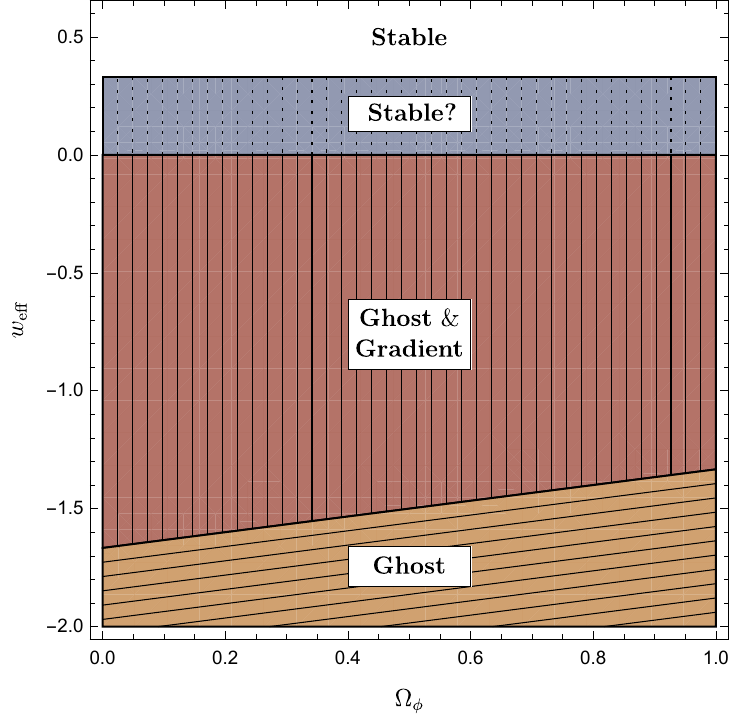}
		\caption{\label{fig:plano weff OmegaPhi} Stability of scalar perturbations in the $(\Omega_\phi,\, w_{\textup{eff}})$-plane for $K\equiv0$. The lower oblique-lined region corresponds to the presence of a the ghost instability. Scalar perturbations are both ghost- and gradient-unstable in the central vertical-lined region. Stability is ensured in the $w_{\textup{eff}}>1/3$ zone, since $w_\phi$ is clearly positive there in virtue of the relation $w_{\textup{eff}}=\Omega_r/3 + w_\phi\Omega_\phi$.
        In the dotted region, which corresponds the horizontal band $0\leq w_{\textup{eff}}\leq1/3$, perturbations may be healthy if and only if $w_{\phi}$ is non-negative, otherwise they will be both ghost- and gradient-unstable. Nevertheless, the sign of $w_\phi$ in that region of the $(\Omega_\phi,\, w_{\textup{eff}})$-plane cannot be inferred without solving the whole evolution of the model and, therefore, the stability conditions there cannot be further addressed without specifying the function $G$.}
    \end{figure}
    
    A similar derivation can also be applied to the
    expression for the speed of sound squared in equation (\ref{eq:gradstab}). Straightforward differentiation over the definition for $\alpha_B$ in equation (\ref{eq:alphaB}) leads to
    \begin{align}
        \frac{\dot{\alpha}_\textrm{B}}{H}=3\left(w_{\textup{eff}}-w_\phi\right)\Omega_\phi,
    \end{align}
    where $w_{\textup{eff}}\coloneqq p_{tot}/\rho_{tot}$ represents the effective equation of state parameter of the total fluid on the r.h.s. of equations (\ref{FE1}) and (\ref{FE2}). Taking this result into account, and also substituting $\dot{H}$ by means of equation (\ref{FE2}), the gradient-free condition (\ref{eq:gradstab}) can be re-formulated in terms of background densities as
    \begin{align}
		c_s^2=\frac{\Omega_\phi(5-\Omega_\phi+3w_{\textup{eff}})}{2D}\geq0.\label{eq:gradstabPB}
    \end{align}
    The numerator in the above expression can be positive or negative depending on the value of $\Omega_\phi$ and $w_{\textup{eff}}$.
    In Figure \ref{fig:plano weff OmegaPhi}, we represent the fulfilment of the stability conditions in the $(\Omega_\phi,\, w_{\textup{eff}})$-plane. As there can be checked, it is possible to have ghost-free and gradient-free scalar perturbations during radiation and matter dominated epoch; that would correspond to $ w_{\textup{eff}}\approx1/3$ and $ w_{\textup{eff}}\approx0$ for $\Omega_\phi\ll 1$, respectively. A necessary and sufficient condition for this to happen is that of $w_\phi$ being positive; see inequality (\ref{eq:ghostPB}).  On the contrary, perturbations will always become unstable at some point if $w_\phi$ is negative; for example, if the braiding scalar field acts as DE. 
    Hence, an accelerating universe cannot be safely modelled when $K\equiv0$ in the KGB action (\ref{eq:actionKGB}). (Recall that acceleration demands the violation of the Strong Energy Condition and, therefore, $w_{\textup{eff}}<-1/3$ at some moment in the recent past. According to Figure \ref{fig:plano weff OmegaPhi}, this would inevitably produce ghost and gradient unstable scalar perturbations.)
    Interestingly, $c_s^2$ could be positive for a super-accelerated regime ($w_{\textup{eff}}<-1$), although this would only be possible at the price of having a ghost mode. This feature may be interesting for modelling phantom DE since, as we have discussed in the previous section, the positivity of the speed of sound squared can be argued to be more fundamental from a classical point of view than the absence of a ghost. However, from Figure \ref{fig:plano weff OmegaPhi} it is clearly not possible to have a braiding model (with $K\equiv0$) that connects this gradient-stable supper-accelerated (late-time) regime with a realistic early-universe description in a way that $c_s^2$ remains always positive. Therefore, modelling DE with only the braiding function $G$ of the KGB theory seems rather unwisely.


\subsection{The complete KGB theory\label{sec:fullKGB}}
In the previous sections we followed a novel approach to the evaluation of the stability of linear scalar perturbations by rewriting the ghost-free (\ref{eq:scalarstab}) and gradient-free (\ref{eq:gradstab}) conditions in terms of background quantities, like the partial energy densities $\Omega_i$ and the equation of state parameters $w_\phi$ and $w_{\textup{eff}}$. Our main purpose with this approach is to have a better physical intuition about the requirements for a given model to produce stable scalar perturbations. \change{From this analysis, we found that instabilities are always triggered when $1+w_\phi<0$ (violation of the NEC in the DE sector) in the case of $k$-essence. Moreover, modelling \textit{kinetic braiding} only with the function $G$ surprisingly makes things even worse, since now instabilities appear as $w_\phi$ becomes negative. Thus, no stable DE component (for which $w_\phi$ should be less than -$1/3$) can be produced in the case of $K\equiv 0$.}
In this section, we will apply the very same analysis now for the full KGB theory.

When both $K$ and $G$ functions are non-trivial, the ghost-free condition (\ref{eq:scalarstab}) cannot be re-expressed solely in terms of the previous background quantities but
\begin{align}\label{eq:ghoststabFKGB}
    D=\frac{3\Omega_J^B\left[3\Omega_J(\Omega_J^B-2)+\left(3+\Omega_r-3\Omega_\phi\right)\Omega_J^B\right]}{6\left[\Omega_J-(1+w_\phi)\Omega_\phi\right]}>0,
	\end{align}
where  
    \begin{align}
	\Omega_J\coloneqq&\,\frac{\dot{\phi}J}{3H^2},\label{def:Omega_J}\\
        \Omega_J^B\coloneqq&\,\frac{2\dot{\phi}XG_X}{H},\label{def:Omega_JB}
    \end{align}
are the partial energy densities for the contributions of the shift-current, $J$, and that of the braiding function, $G$, to the different terms in the expression for scalar field energy density $\rho_\phi$ in equation (\ref{rho_phi}), respectively.  Please note that $\Omega_J^B$ coincide with the definition of $\alpha_{\rm B}$ \cite{Bellini:2014fua}, see equation (\ref{eq:alphaB}).
It is also important to highlight that the physical constraint of $\Omega_\phi\in[0,1]$ does not imply, in general, any bound on the value of these auxiliary partial energy densities. Their value can be arbitrary large and/or negative. 

Similarly, the gradient-free condition (\ref{eq:gradstab}) can be translated to 
    \begin{align}\label{eq:gradstabFKGB}
		c_s^2=\frac1D\left\lbrace\frac{\Omega_J^B(2-\Omega_J^B)}{2}+3\Omega_J+3(2+\eta)\left[\Omega_J-(1+w_\phi)\Omega_\phi\right]\right\rbrace\geq0,
    \end{align}	
being
    \begin{align}\label{def:eta}
		\eta\coloneqq\frac{2XG_{XX}}{G_X},
    \end{align}
a relevant quantity for the discussion in the next section. Please note that when $G$ or $K$ are trivial, the above expressions reduce to those obtained in sections \ref{sec:kessence} and \ref{sec:PB}, respectively. 

Contrary to the limiting cases analysed in the previous sections, conditions (\ref{eq:ghoststabFKGB}) and (\ref{eq:gradstabFKGB}) can be satisfied simultaneously for a wide variety of models including those with phantom behaviour. For instance, considering $\eta$ to be positive (since this will be crucial in the incoming sections) we show all the possible cases where both inequalities are satisfied in  \ref{app:FKGB}. 
Nevertheless, since $\Omega_J^B$ and $\Omega_J$ are not physically meaningful quantities as $\Omega_\phi$ is, it is difficult to extract the most general conditions the functions $K$ and $G$ should meet in order to produce a stable model. Anyway, there are examples of such models in the literature. For example, in reference \cite{KGB1} the authors analyse the case of $K$ and $G$ being proportional to $X$. They also show there that the resulting cosmological model has completely stable linear scalar perturbations even though the scalar field is violating the NEC. (Note, however, that their model does not feature super-acceleration since $1+w_{\textup{eff}}>0$ always.)
More examples of stable KGB models can be found in references \cite{DeFelice:2011bh,DeFelice:2011aa,Giacomello:2018jfi,Frusciante:2019puu}. Hence, it is possible to construct reliable cosmological models with the scalar field coming from shift-symmetric KGB theories when both $k$-essence and braiding functions are non-trivial.


\subsection{Summary}
Could we describe a stable DE component with a scalar field coming from shift-symmetric KGB theories?
There is no problem to describe it with a $k$-essence term if it satisfies the NEC. However, it is well-known that $k$-essence suffers from instabilities if the NEC is violated  \cite{Hsu:2004vr,Abramo:2005be,Creminelli:2008wc,Rubakov:2014jja}. Moreover, the scalar field energy density should be always phantom or non-phantom since $k$-essence cannot produce a phantom-crossing \cite{Vikman:2004dc} (see also \cite{Caldwell:2005ai}). Hence, if we interpret current data as favouring the presence of a slight violation of the NEC in the DE sector at present times, then, the scalar field should have always violated the NEC. In this scenario, a $k$-essence proposal will be plagued with a gradient or a ghost instability. With an appropriate selection for the function $K$, however, the gradient-free condition can be satisfied for (even) NEC-violating models. Nevertheless, the resulting scalar field  will feature a ghost DOF; yet this may not always be catastrophic. As we have discussed, it would be interesting to explore whether it is possible to have a ghost mode in cosmology without destabilising the classical nor the semi-classical regimes of the theory. While this may be feasible, addressing both the $H_0$ and $S_8$ tensions (in case the latter is \change{still present in future surveys}) by means of late-time physics only continues to pose an insurmountable obstacle in the $k$-essence scenario, since no phantom-crossing is possible (see discussion in references \cite{Heisenberg:2022lob,Heisenberg:2022gqk}).

The situation is significantly worsen when only the braiding function $G$ is present in the scalar field's action (\ref{eq:actionKGB}). Scalar perturbations are always ghostly if the scalar field acts as a DE component, i.e. if $w_\phi<-1/3$ (see condition (\ref{eq:ghostPB})). This is aggravated even further by the fact that a gradient instability is also present in any realistic model that interpolates between an initial radiation dominated epoch and a late-time accelerated expansion regime; see Figure \ref{fig:plano weff OmegaPhi}. Hence, the term $G\Box \phi$ alone in the action (\ref{eq:actionKGB}) is not a viable option for modelling DE.

Finally, when considering the sum of $k$-essence and the braiding term the situation is different. Following the same strategy we have used to analyse the preceding marginal examples, we have recovered the well-known result that the conditions for the avoidance of ghost and gradient instabilities can be simultaneously fulfilled even when the $\phi$-fluid violates the NEC (see  \ref{app:FKGB}). Although the lack of a physical intuition for the auxiliary quantities $\Omega_J$ and $\Omega_J^B$ did not allow us to restrict the possible phenomenology of the theory as stringently as for the previous cases, note that linearly stable KGB models have been already reported in the literature (see for instance \cite{DeFelice:2011bh,DeFelice:2011aa,Giacomello:2018jfi,Frusciante:2019puu} and references therein). This fact points to the viability (from the point of view of the stability of linear scalar perturbations) of modelling a realistic DE component with the scalar field from the (complete) KGB theory, at least in principle.


\section{Beyond scalar perturbations\label{sec:intGWs}}

In section \ref{sec:linearPerturbs} we have discussed the absence of ghost and gradient instabilities in the linearised scalar perturbations around a flat FLRW background. In that approach, interactions between scalar, vector and tensor perturbations were not present, since they decoupled at the linear level. However, these interactions can appear when going beyond the linear regime and may be important. In fact, the interaction between tensor perturbations (essentially GWs) and DE fluctuations may induce ghost and/or gradient instabilities in the scalar sector, as it was pointed out in reference \cite{Creminelli3}. In that case, the stability conditions discussed in the previous sections do not guarantee the viability of the theory\footnote{See also references \cite{Creminelli1,Creminelli2} for a discussion on the decay of GWs into scalar fluctuations when Lorentz invariance is spontaneously broken.}, since the braiding term in the action can destabilize the theory through the GWs-DE interaction. As a concrete example, for cubic Galileon (which corresponds to $G(X)\propto X$) this effect is detrimental when $|\alpha_B|\geq10^{-2}$; that is to say, if the effects of the braiding are sizeable \cite{Creminelli3}.

In general, the result of reference \cite{Creminelli3} raises serious doubts on the viability of any KGB theory (different from $k$-essence) aiming to describe the currently observed abundance of DE in the Universe. Nevertheless, as the authors of that reference commented (see also reference \cite{Zumalacarregui:2020cjh}), it is also important to analyse the fate of the instabilities once they are originated. The fate of the theory after the instability is reached is, however, uncertain since it depends on an unknown UV-completion.

In the following, we review the arguments and notation presented in reference \cite{Creminelli3}; that is section \ref{sec:revGWs}.
\change{In section \ref{sec:noGrad} we obtain the general conditions a KGB model should meet in order to not be destabilised when interacting with GWs, even when the contribution from the braiding is non-negligible. The latter possibility turns to be attainable only if the braiding function $G$ is not linear in the kinetic term $X$ (i.e. if it is different from that of cubic Galileon).}
In section \ref{sec:noExample} we explore whether these conditions could be satisfied by some reasonable KGB model.


\subsection{Interaction with gravitational waves: a review \label{sec:revGWs}}

In reference \cite{Creminelli3} the authors use the EFT approach to discuss the interaction between DE perturbations and GWs. (See  \ref{app:EFT} for a review on the EFT approach to DE.) For this effect to be sizeable one needs a cubic coupling $\pi\pi\gamma$ in the perturbed action, where $\gamma$ stands for tensor perturbations (essentially GWs) and $\pi$ denotes scalar field fluctuations  \cite{Creminelli3} (not to be confused with the scalar field $\phi$ itself). In the covariant version of the theory, the interaction $\pi\pi\gamma$ is related to the Cubic Horndeski operator; i.e.  the braiding term $G\Box\phi$ in action (\ref{eq:actionKGB}). Nevertheless, this interaction vertex is also present in beyond Horndeski theories \cite{Creminelli3,Creminelli1,Creminelli2}.

The interaction between DE perturbations and GWs is conveniently addressed in the Newtonian gauge \cite{Creminelli3}
\begin{align}\label{eq:NewtonianG}
    g_{00}=-(1+2\Phi) \ \ \ \textup{and} \ \ \ g_{ij}=a^2(1-2\Psi) (e^\gamma)_{ij},
\end{align}
where $\Phi$ and $\Psi$ are the Newtonian potentials, and $\gamma_{ij}$ is transverse and traceless.
At leading order, the Lagrangian density describing this interaction reads\footnote{This Lagrangian density follows from restoring time diffeomorphism in the EFT action (\ref{actionEFT_KGB}) for the KGB theory using the St\"uckelberg procedure. That is, by performing in the action (\ref{actionEFT_KGB}) the transformation $t\to t+\pi(t,x^i)$, where $\pi$ denotes the scalar field fluctuations. The relevant transformations in this case are \cite{Cusin:2017mzw} (see also \cite{Creminelli3})
\begin{align}
    \delta g^{00}\to& \, \delta g^{00}+2g^{0\mu}\partial_\mu\pi+g^{\mu\nu}\partial_\mu\pi\partial_\nu\pi, \label{eq:delta g00}\\
    \delta K\to&\, \delta K-(1-\dot\pi) g^{ij}\partial_i\partial_j\pi+\frac{2}{a^2}\partial_i\pi\partial^i\dot\pi+\dots\,,
\end{align}
where the former transformation is an exact relation to all order in $\pi$, whereas the latter has been expanded up-to second order in perturbations. 
After restoring full gauge invariance, the Newtonian gauge (\ref{eq:NewtonianG}) can be implemented. 
The Lagrangian density (\ref{eq:Lpi}) is finally obtained after solving (at linear order) for the Newtonian potentials in terms of $\pi$, and canonically normalizing  the scalar and tensor fluctuations as \cite{Creminelli3}
\begin{align}
    \pi\to\,\sqrt{D}H\pi\ \ \ \ \rm{and}\ \ \ \
    \gamma_{ij}\to\, \frac{1}{\sqrt{2}}\gamma_{ij},
\end{align}
respectively, being $D$ the ghost function (\ref{eq:scalarstab}).
} 
\cite{Creminelli3}
\begin{align}\label{eq:Lpi}
\mathcal L_\pi =& \frac{1}{2}\left[ \dot{\pi}^2 - {c_s^2}\,\partial_i\pi\partial^i\pi \right] -\frac{1}{\Lambda_{\rm B}^3} \square \pi \partial_\mu \pi \partial^\mu\pi+ \frac{\eta}{\Lambda_{\rm B}^3}\ddot \pi \partial_i \pi\partial^i \pi \nonumber\\
&+ \Gamma^{\mu \nu} \partial_{\mu} \pi \partial_\nu \pi - \frac{\Lambda_{\rm B}^3}{2}\pi\Gamma_{\mu \nu} \Gamma ^{\mu \nu},
\end{align}
where the expansion of the Universe has been neglected and indices are raised/lowered with the Minkowski metric. In addition, $c_s^2$ is given by expression (\ref{eq:gradstab}) and 
\begin{align}
    \Gamma_{\mu \nu} &\coloneqq  \frac{\dot \gamma_{\mu \nu} }{ \Lambda^{2}},\label{def:Gamma}\\
    \Lambda^2&\coloneqq \frac{\sqrt2 H^2 D}{\bar{m}_1^3},\\
    \Lambda_{\rm B}^3&\coloneqq -\frac{2H^3 D^\frac32}{\bar{m}_1^3},
\end{align}
being $D$ the ghost factor (\ref{eq:scalarstab}). Moreover, the parameter $\eta$ is defined through the relation \cite{Creminelli3}
\begin{align}\label{etaEFT}
4\bar{m}_2^3=-(1+\eta)\bar{m}_1^3,
\end{align}
where $\bar{m}_1^3$ and $\bar{m}_2^3$ are functions on the background time, $t$, that appear in the EFT action (\ref{actionEFT_KGB}).
This parameter is useful for measuring deviations with respect to cubic Galileon (i.e. $G(X)\propto X$) for which $\eta$ is zero and, therefore, $4\bar{m}_2^3=-\bar{m}_1^3$. (Further details can be found in  \ref{app:EFT}.) The covariant version of this parameter can be obtained from the expression for the mass parameters $\bar{m}_1^3$ and $\bar{m}_2^3$ in terms of the function $G$ and its derivatives (see  \ref{app:EFT}). Taking the expressions (\ref{eq:m13}) and (\ref{eq:m23}) into account, it is straightforward to obtain that the covariant version of $\eta$ is nothing but the quantity we already have defined in equation (\ref{def:eta}). That is $\eta=2XG_{XX}/G_X$.

The field equation for the scalar field perturbation, $\pi$, follows from varying the action corresponding to the Lagrangian density (\ref{eq:Lpi}). This reads \cite{Creminelli3}
\begin{align}\label{eom:pi}
    &\ddot\pi-c_s^2 \nabla^2\pi+\frac{2}{\Lambda^3_{\rm B}}\left[(\partial_\mu\partial_\nu\pi)(\partial^{\,\mu}\partial^\nu \pi)-\left(\Box\pi\right)^2\right]+\frac12 \Lambda_{\rm B}^3\Gamma_{\mu\nu}\Gamma^{\,\mu\nu}\nonumber\\
    &+\frac{2\eta}{\Lambda_{\rm B}^3}\left[(\partial_i\dot\pi)(\partial^i\dot\pi)-\ddot\pi \nabla^2\pi\right]+2\Gamma^{\,\mu\nu}\partial_\mu\pi\partial_\nu\pi=0,
\end{align}
where $\nabla^2\equiv\eta^{ij}\partial_i\partial_j$. 
Here we will restrict our discussion to the case where $\pi$ propagates  subluminally, that is when $c_s^2<1$ \cite{Creminelli3}. (The luminal and superluminal cases are discussed in reference \cite{Creminelli3}.) In this scenario, the lightcone for $\pi$ is narrower than the one for the GWs, since $c^2_{\rm GWs}=1$ \cite{Creminelli3,Creminelli2}. Consequently, the DE fluctuations are not sensitive to the source of the GWs, provided we are far enough from the source of emission \cite{Creminelli3,Creminelli2}. 
To address the DE-GWs interaction, therefore, we can consider the following classical GW solution with linear polarization ``+'' travelling in the $z$ direction \cite{Creminelli3,Creminelli2}
\begin{align}\label{eq:gamma}
\gamma_{ij} =  h_0^+ \sin \left[ \omega (t-z) \right] \epsilon_{ij}^+ ,
\end{align}
being $\epsilon_{ij}^+ = \text{diag}(1,-1,0)$ and $h_0^+$  the dimensionless strain amplitude. Then, it follows from the definition (\ref{def:Gamma})  that
\begin{align}
    \label{Gamma}
    \Gamma_{00} = \Gamma_{0i} =0 \;, \qquad \Gamma_{ij} = \frac{\beta c_s^2}{2} \cos \left[ \omega (t - z) \right] \epsilon_{ij}^+,
\end{align}
being
\begin{align} \label{beta}
\beta \coloneqq \frac{2 \omega  h_0^+}{c_s^2 |\Lambda^2|} = \frac{\sqrt{2}  \omega h_0^+}{H^2 c_s^2} \left|\frac{\bar{m}_1^3}{D}\right|,
\end{align}
a parameter. 

The field equation (\ref{eom:pi}) simplifies significantly when written in the null coordinates \cite{Creminelli3}
\begin{align}
    u\coloneqq t-z \ \ \ \textup{and} \ \ \ v\coloneqq t+z.
\end{align}
Since there is no intersection between the region where the source of emission of the GWs is active and the past lightcone of $\pi$, there is a translational invariance along the latter null coordinate \cite{Creminelli3}. This suggests the \"anstaz  $\pi=\pi(u)$ for the solutions of the field equation (\ref{eom:pi}).
By defining
\begin{align}
    \varphi\coloneqq \frac{\pi}{\Lambda_{\rm B}^3},
\end{align}
equation (\ref{eom:pi}) reduces to \cite{Creminelli3}
\begin{align}\label{eom:varphi}
    \varphi''(u)=-\frac{\Gamma_{\mu\nu}\Gamma^{\,\mu\nu}}{2(1-c_s^2)}=-\frac{\beta^2 c_s^4}{4(1-c_s^2)}\cos^2(\omega u),
\end{align}
where prime denotes derivation w.r.t. the null coordinate $u$. In addition, the relations (\ref{Gamma}) have been used in the second equality. From this result it follows that $\varphi$ is always decelerating along the null direction $u$.

The stability of a general solution to the equation (\ref{eom:varphi}), namely $\hat\pi$, can be addressed  
by analysing the tensor structure of the kinetic part related to the quadratic action for small fluctuations around that solution.
That is, by considering $\pi=\hat\pi+\delta\pi$ and expanding the Lagrangian density (\ref{eq:Lpi}) around the background solution $\hat{\pi}$ coming from equation (\ref{eom:varphi}). This leads to \cite{Creminelli3} (see also reference \cite{Nicolis:2004qq})
\begin{align}\label{eq:Ldpi}
\mathcal{L}^{(2)}_{\rm kinetic}=Z^{\mu\nu}(\hat\pi)\,\partial_\mu\delta\pi\,\partial_\nu\delta\pi,
\end{align}
where the values of the kinetic matrix $Z^{\mu\nu}$ depend on the space-time coordinates through the background solution $\hat\pi$. 
These are \cite{Creminelli3}
\begin{subequations}\label{eq:Zs}
\begin{align}
Z^{00} &= \frac12 +(2+\eta)\varphi''(u)\nonumber\\
&=\frac{1}{2}\left[1-\frac{c_s^4\beta^2}{2(1-c_s^2)}(2+\eta)\cos^2
\left(\omega u\right)\right], \label{eq:Z00}\\
Z^{11} &= - \frac{c_s^2}{2}+\Gamma^{11}+\eta\varphi''(u)\nonumber\\
&=- \frac{c_s^2}{2}\left[ 1-\beta\cos \left(\omega u\right)+\frac{c_s^2\beta^2}{2(1-c_s^2)}\eta\cos^2 \left(\omega u\right)\right] , \label{eq:Z11}\\
Z^{22} &= - \frac{c_s^2}{2}+\Gamma^{22}+\eta\varphi''(u)\nonumber\\
&=- \frac{c_s^2}{2}\left[ 1+\beta\cos \left(\omega u\right)+\frac{c_s^2\beta^2}{2(1-c_s^2)}\eta\cos^2 \left(\omega u\right)\right] , \label{eq:Z22}\\
Z^{33} &=- \frac{c_s^2}{2}+(2+\eta)\varphi''(u)\nonumber\\
&=-\frac{c_s^2}{2}\left[1+\frac{c_s^2\beta^4}{2(1-c_s^2)}(2+\eta)\cos^2\left(\omega u\right)\right], \\
Z^{03} &= Z^{30} =(2+\eta)\varphi''(u)\nonumber\\
&= -\frac{c_s^4\beta^2}{4(1-c_s^2)}( 2 + \eta) \cos^2 \left(\omega u\right),
\end{align}
\end{subequations}
where $\varphi''$ is given by equation (\ref{eom:varphi}). (Recall that we are considering here only the case of subluminal propagation for $\hat\pi$, i.e. $c_s^2<1$. See reference \cite{Creminelli3} for a discussion when $c_s^2\geq1$.)

It should be highlighted that for a general time-dependent kinetic term like (\ref{eq:Ldpi}) there is no clear definition of stability \cite{Nicolis:2004qq}. However, in the limit where the time and length scales considered are much shorter than the rate of variation of $\hat\pi$, it is perfectly acceptable to analyse the stability of the system as if the kinetic matrix $Z^{\mu\nu}$ were constant \cite{Nicolis:2004qq} (see also reference \cite{Creminelli3}).
Within this \emph{local} approximation to the stability of the theory, the absence of a gradient instability will be determined by the correct relative signs between the terms involving time and spatial derivatives in the Lagrangian density (\ref{eq:Ldpi}). It is convenient, however, to move to Fourier space and directly analyse the dispersion relation for a mode with four-momentum $k^\mu=(\omega,k^i)$ since the kinetic matrix $Z^{\mu\nu}$ is not in diagonal form. For the Lagrangian density (\ref{eq:Ldpi}), the dispersion relation reads \cite{Nicolis:2004qq}
\begin{align}
    Z^{\mu\nu} k_\mu k_\nu=0.
\end{align}
Moreover, multiplying this relation by $Z^{00}$, it is possible to re-expressed it as \cite{Nicolis:2004qq}
\begin{align}
    \left(Z^{0i}k_i- Z^{00}\omega\right)^2=\left(Z^{0i}Z^{0j}-Z^{00}Z^{ij}\right)k_i k_j,
\end{align}
when $Z^{00}$ is not trivial. We recall that indices have been raised/lowered with the Minkowski metric.
The absence of a gradient instability demands all frequencies, $\omega$, to be real for any wave vector $k^i$. Otherwise, exponentially growing modes will be present. It is straightforward to notice from the above expression that this stability is guaranteed if the spatial matrix $ Z^{0i}Z^{0j}-Z^{00}Z^{ij}$ is positive defined \cite{Nicolis:2004qq}. Since this matrix is diagonal for the $Z^{\mu\nu}$ introduced in equations (\ref{eq:Zs}), in our case of interest the positive-definedness condition reduces to
\begin{align}\label{eq:gradZ}
   Z^{0i}Z^{0j}-Z^{00}Z^{ij}\geq0,
\end{align}
for all spatial directions. 

In addition to the previous condition, the ghost-like instability is avoided if \cite{Nicolis:2004qq}
\begin{align}\label{eq:ghostZ}
    Z^{00}>0.
\end{align}
It is interesting to note that a theory with $Z^{00}<0$ can be also made stable provided it features superluminal excitations and that one can boost to a frame with $Z^{00}>0$ \cite{Dubovsky:2005xd}. However, we will not discuss this case here.
Also note that  the conditions (\ref{eq:gradZ}) and (\ref{eq:ghostZ}) reduce to the standard ones when the matrix $Z^{\mu\nu}$ is in diagonal form \cite{Rubakov:2014jja}.

When $\eta$ is always trivial (that is when considering the case of cubic Galileon only, i.e. $G\propto X$), the ghost-free condition (\ref{eq:ghostZ}) leads to
\begin{align}\label{cond:noghost eta0}
    |\,\beta\,|< \frac{1}{c_s^2}\sqrt{1-c_s^2}\,,
\end{align}
which follows from demanding the factor ahead of the trigonometric term in equation (\ref{eq:Z00}) to be less than one. Moreover, the expression $(Z^{03})^2-Z^{00}Z^{33}$ is always positive when $\eta\equiv0$. Therefore, there is not a gradient instability in the $z$ direction.
The background solution $\hat\pi$, on the other hand, does not affect the entries $Z^{11}$ and $Z^{22}$. For $|\,\beta\,|>1$, these entries  oscillate between positive and negative values in a way that condition (\ref{eq:gradZ}) could never be fulfilled. Avoiding this oscillations in the $(x,y)$-plane demands
\begin{align}\label{cond: beta eta0}
    |\,\beta\,|<1.
\end{align}
For a non-relativistic speed of sound, that is $c_s\ll1$, condition (\ref{cond: beta eta0}) is clearly more restrictive than the ghost-free condition (\ref{cond:noghost eta0}). Therefore, by demanding $|\,\beta\,|<1$ the stability conditions (\ref{eq:gradZ}) and (\ref{eq:ghostZ}) are simultaneously fulfilled in the non-relativistic regime. 

The  complete stability region (including relativistic $c_s$) is shown in green in figure \ref{fig:stabGWs}. As there can be seen, $|\,\beta\,|<1$ is a necessary condition for the absence of ghost and gradient instabilities when $\eta$ is trivial\footnote{Note, however, that this conditions is not sufficient to guarantee the stability for values of $c_s$ close to the speed of light.}.
Owing to the definition of $\beta$ in equation (\ref{beta}), this condition results in a tight constraint over $\bar{m}_1^3$ \cite{Creminelli3}. Please note that there is a direct relation between $\bar{m}_1^3$ and $\alpha_{\textrm{B}}$, which essentially measures the amount of \textit{braiding} in the theory. This is 
\begin{align}
    \bar{m}_1^3=\alpha_{\rm B},
\end{align}
as follows from comparing equations (\ref{eq:alphaB}) and (\ref{eq:m13}). 
The observational bounds for $\alpha_{\rm B}$ coming from the stability condition of $|\,\beta\,|<1$ are discussed in reference \cite{Creminelli3}. To that aim, the authors there have considered a GWs-background as that generated by the event  GW150914 \cite{LIGOScientific:2016aoc} with a chirp mass of $M_c = 28M_\odot$ and $f = 30$ Hz at a distance of 1 Mpc. They found that the condition of $|\,\beta\,|<1$ directly translates to $|\alpha_{\textrm{B}}|\leq 10^{-2}$ for that event \cite{Creminelli3}, see definition of $\beta$ in equation (\ref{beta}). 
Thus, they conclude that the braiding part should not have any sizeable effect in order to avoid the destabilising interaction with GWs when $\eta$ is zero \cite{Creminelli3}. Similar conclusions apply also to the case of $c_s=1$ \cite{Creminelli3}. For the superluminal case, $c_s>1$, it has been argued that the same results should hold as well, but no explicit calculations were provided due to technical difficulties \cite{Creminelli3}.

On the contrary, if $\eta\neq0$, then, the oscillations in the spatial entries $Z^{11}$ and $Z^{22}$ can be avoided even for $|\,\beta\,|>1$
provided that $\eta$ is \textit{sufficiently} large and positive (see appendix  A of reference \cite{Creminelli3}). Hence, this may provide a mechanism for having a stable theory with $|\,\beta\,|>1$. If that is the case, then, the strong observational bounds for $\alpha_{\rm B}$ discussed in reference \cite{Creminelli3} shall be revised.
The larger the value of $\eta$, however, the harder avoiding a ghost instability will be since this quantity enters linearly in the factor ahead of the trigonometric term in expression (\ref{eq:Z00}). 
Therefore, one should carefully explore the region of viability for the optimal effect when $\eta>0$.
We address this discussion in the next sections.


\subsection{Loophole hunting\label{sec:noGrad}}

Condition (\ref{cond: beta eta0}) places a strong restriction to the maximum amount of braiding a theory may have in order to avoid the destabilising interaction with GWs described in reference \cite{Creminelli3}. As a result from that constraint, the braiding part should not have any sizeable effects when data from GWs events are taken into account \cite{Creminelli3}.
However, this restrictions holds only for the special case of $\eta\equiv0$.  In a more general situation, it would be possible to ease the constraint in equation (\ref{cond: beta eta0}) provided that $\eta$ is positive enough to ensure that all spatial direction in the kinetic matrix $Z^{\mu\nu}$ are negative. 
This is obviously not the case for $G\propto X$ for which $\eta$ is trivial \cite{Creminelli3}, but may be possible for models beyond a linear function $G$. 

Inspired by the fact that the ghost-free condition (\ref{cond:noghost eta0}) is less restrictive than the constraint in equation (\ref{cond: beta eta0}) for non-relativistic sound velocities when $\eta$ is trivial, we explore here the possibility of a ghost-free and gradient-free theory with $|\,\beta\,|>1$ when $\eta\neq0$. 
If that possibility comes to be true, then, the previously mentioned  observational bound of $|\alpha_{\rm B}|\leq 10^{-2}$ \cite{Creminelli3} should be revised. (Recall the definition of $\beta$ in equation (\ref{beta}).)  This may indicate that KGB theory could have sizeable effects at cosmological scales after all. 

In general, the ghost-free condition (\ref{eq:ghostZ}) implies
\begin{align}\label{cond:noghost etaNo0}
    |\,\beta\,|<\frac{1}{c_s^2}\sqrt{\frac{2(1-c_s^2)}{2+\eta}},
\end{align}
which follows from demanding the factor ahead of the trigonometric term in equation (\ref{eq:Z00}) to be less than one. The gradient-free condition (\ref{eq:gradZ}) is automatically satisfied in the $z$ direction since the combination
\begin{align}
    (Z^{03})^2-Z^{00}Z^{33}=\frac{c_s^2}{4}\left[1+\frac{(2+\eta)\beta^2c_s^2}{2}\cos^2(\omega u)\right],
\end{align}
is always positive for the values of interest for $\eta$. Whereas in the $xy$-plane the absence of a gradient instability implies
\begin{align}
    Z^{00}\,Z^{11}<0,\label{cond:gradZ11}\\
    Z^{00}\,Z^{22}<0,\label{cond:gradZ22}
\end{align}
as follows from comparing the gradient-free condition (\ref{eq:gradZ}) with the expressions for the $Z^{\mu\nu}$ found in equations (\ref{eq:Zs}).
Contrary to the case of $\eta\equiv0$, the oscillations in the spatial entries $Z^{11}$ and $Z^{22}$ can be avoided even for $|\,\beta\,|>1$
provided that $\eta$ is \textit{sufficiently} large and positive. Taking into account their explicit expressions  in equations (\ref{eq:Z11}) and (\ref{eq:Z22}), respectively, we found that for
\begin{align}\label{eq:noGWsGrad}
    \eta> \frac{1-c_s^2}{2c_s^2},
\end{align}
these terms are always negative. A model satisfying the above condition will feature positive entries $Z^{11}$ and $Z^{22}$. Therefore, it will be free from gradient instabilities in the $xy$-plane provided that $Z^{00}$ is positive. That is to say, if the ghost-free condition (\ref{cond:noghost etaNo0}) is also satisfied. In other words, condition (\ref{cond:noghost etaNo0}) and the new constraint found in equation (\ref{eq:noGWsGrad}) ensure the total stability of the system.

\begin{figure}
    \centering
    \includegraphics[width=\columnwidth]{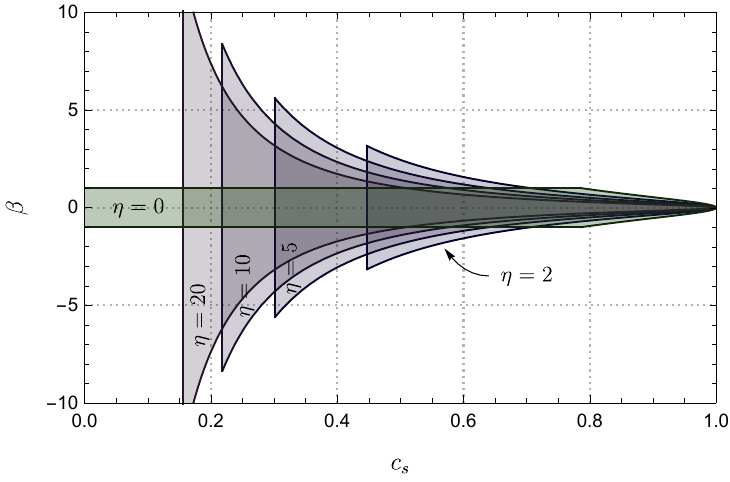}
    \caption{Slices at constant $\eta$ of the stability region where conditions (\ref{cond:noghost etaNo0}) and (\ref{eq:noGWsGrad}) are simultaneously satisfied. Within these regions, the interaction between DE fluctuations and GWs described in reference \cite{Creminelli3} does not induced a ghost or gradient instability. The green zone represent the stability at $\eta=0$. This is the corresponding region for cubic Galileon, i.e. $G\propto X$ (compare with figure 2 in reference \cite{Creminelli3}). The darker shaded areas represent the stability at different non-zero values of $\eta$.}
    \label{fig:stabGWs}
\end{figure}

We should emphasise the fact that condition (\ref{eq:noGWsGrad}) does not depend on the specific properties of the GWs (local properties) but only on the value of $c_s^2$, which depends on the KGB functions. It is also interesting to note that if scalar perturbations propagate close to the speed of light, then, condition (\ref{eq:noGWsGrad}) is easily satisfied if $\eta$ is just slightly positive. However, this regime would strongly constraint $\beta$ to vanish, see equation (\ref{cond:noghost etaNo0}). For a model with a non-negligible amount of braiding, this suggests that avoiding GWs-induced instabilities becomes more difficult if DE fluctuations propagate with a relativistic speed of sound. 
Conversely, the ghost-free condition (\ref{cond:noghost etaNo0}) is compatible with $|\,\beta\,|\gg1$ if $c_s^2$ is small. Nevertheless, $\eta$ should diverge strongly than $1/c_s^2$ in order to avoid the gradient instability in this regime. 
This behaviour is summarised in figure \ref{fig:stabGWs}. There we show different slices at constant $\eta$ of the stability region where conditions (\ref{cond:noghost etaNo0}) and (\ref{eq:noGWsGrad}) are simultaneously fulfilled. The green region corresponds to the case of $\eta\equiv0$ (compare with figure 2 in reference \cite{Creminelli3}). As discussed before, the absolute value of $\beta$ is always confined there to be less than one. Alternatively, this region  becomes significantly bigger for small speed of sound if $\eta$ is positive. In fact, stability at $|\,\beta\,|\approx 10$ is possible for $\eta$ of the order of twenty. Please note that this suppose an enlargement of the stability region by a factor of ten w.r.t. the green zone. Moreover, this region can be amplified even more at low $c_s$ for larger values of $\eta$. In the high $c_s$ regime, however, the stability zone narrows compared to the case of $\eta$ trivial.

In view of these results, we postulate that the would-be KGB candidate that could ease (to some extent) the restrictions from  the interaction with GWs described in reference \cite{Creminelli3} must meet the following necessary (but not sufficient) conditions. It should produce a non-relativistic speed of sound, $c_s$, and a large value of $\eta$ \change{ during the expansion history where interactions with GWs are expected}. The definition of both quantities in terms of the KGB functions and its derivatives can be found in equations (\ref{eq:gradstab}) and (\ref{def:eta}), respectively. However, if for the selected model DE perturbations propagate at relativistic speeds, then, smaller values of $\eta$ are preferred. In any case, the contribution of the braiding part in that regime should be negligible.

It should be also mentioned that, within the spirit of the discussion in past sections, one may consider the possibility to relax the ghost-free condition (\ref{cond:noghost etaNo0}) and be concerned only with ensuring a gradient-free spectrum. Of course, this will drastically reduce the restrictions on the theory. Although the possibility of a ghost-mode in cosmology sets an interesting discussion, here there is a technical difficulty in pursuing that direction. Given the structure of $Z^{00}$ in equation (\ref{eq:Z00}), if the condition (\ref{cond:noghost etaNo0}) is violated then the sign of $Z^{00}$ is not always negative (ghost mode excited) but fluctuates. Moreover, this fluctuation will jeopardise the relations (\ref{cond:gradZ11}) and (\ref{cond:gradZ22}) since the functional dependence of $Z^{00}$, $Z^{11}$ and $Z^{22}$ on $\cos(\omega u)$ is different. Thus, if the ghost-free condition is violated, then, gradient instabilities in the $xy$-plane will also (periodically) appear within each fluctuation of the GWs.


\subsection{No natural dark energy\label{sec:noExample}}

A natural KGB candidate to test our hypothesis is that when $\eta$ is a constant. From its expression in terms of $G$ and its derivatives (\ref{def:eta}), this implies that
\begin{align}\label{eq:cteETA}
    G(X)=c_G X^{\frac{2+\eta}{2}},
\end{align}
being $c_G$ a coupling constant. That is a power-law prescription for the braiding function $G$. Consequently, the case of $\eta$ constant provide not only a natural choice to test our proposal but a physically relevant model.
For this model, the resulting speed of sound squared should be constrained as
\begin{align}
    \frac{1}{1+2\eta}<c_s^2<1,
\end{align}
in order to fulfil condition (\ref{eq:noGWsGrad}). This results in a lower limit for $c_s^2$ that should be satisfied (at least) during the cosmological evolution where interactions with GWs are expected. Please note that this is in agreement with the results presented in reference \cite{Albarran:2020bwn}, where the possibility of a vanishing speed of sound was shown to be disfavoured.
It should be also highlighted that, even though having a constant $\eta$ is a very restrictive consideration, the results from exploring this case should also hold (at least asymptotically) for any other model for which $\eta$ converges to a constant value. In other words, the physical conclusions from exploring a model like (\ref{eq:cteETA}) should also apply asymptotically to any KGB theory for which the function $G$ converges, at some point, to a power-law function.

\begin{figure*}[ht]
    \begin{subfigure}{0.4\textwidth}
    \begin{adjustwidth}{-.5cm}{}
    \includegraphics[scale=1.0]{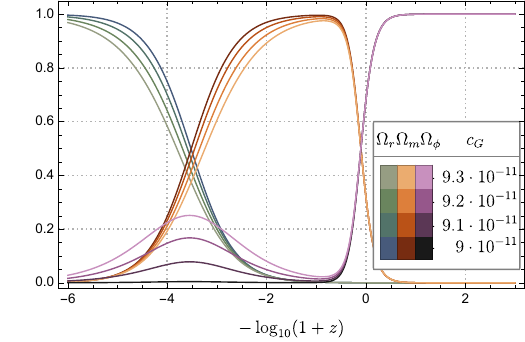}
    \end{adjustwidth}
    \end{subfigure}
    \hfill
    \begin{subfigure}{0.4\textwidth}
    \begin{adjustwidth}{-1.9cm}{}
    \includegraphics[scale=0.725]{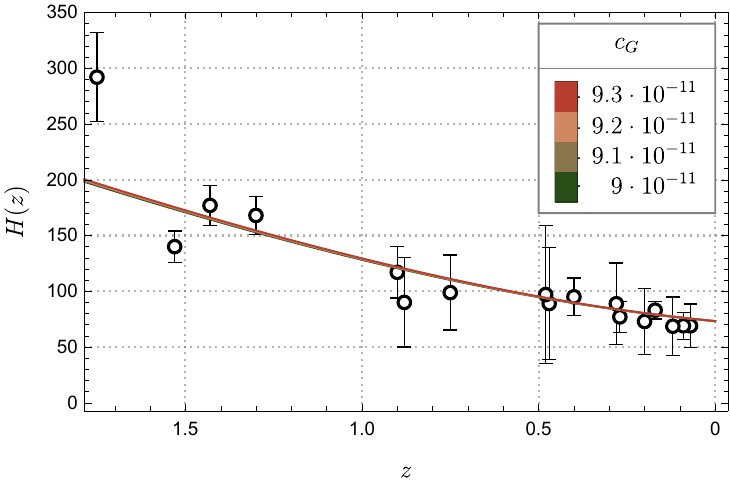}
    \end{adjustwidth}
    \end{subfigure}
\caption{Numerical evolution for $\eta=0.4$ and  $\alpha=0.6$ in expressions (\ref{eq:cteETA}) and (\ref{eq:K}), respectively. In addition, the coupling constant $c_K$ equals minus unity, whereas $c_G$ ranges from $9\cdot10^{-11}$ to $9.3\cdot10^{-11}$. Left panel represents the evolution of the partial energy densities for radiation, matter and the scalar field. Right panels show the evolution of the Hubble rate at low redshift compared to direct measurements collected from references \cite{2014RAA,Simon:2004tf,Ratsimbazafy:2017vga,2010JCAP,Borghi:2021rft} (see also Table II in reference \cite{Alonso-Lopez:2023hkx}).\label{fig:mod_1}}
\end{figure*}

In order to explore whether this mechanism can be conducted by a reasonable KGB model, we select a power-law function for the $k$-essence part as well. That is, we also consider 
\begin{align}\label{eq:K}
    K(X)=c_K X^\alpha,
\end{align}
where $\alpha$ is a constant and $c_K$ equals to $+1$ or $-1$. 
For the KGB model given by the functions (\ref{eq:cteETA}) and (\ref{eq:K}), we have numerically explored the parameter-space spanned by $\{(\eta>0)\times\alpha\times c_G\}$. As our main purpose, at this stage, is to examine the stability of the KGB theory, not to confront it with observational data, we have considered the initial conditions $H_0=73.2$ km s$^{-1}$ Mpc$^{-1}$, $\Omega_{\phi0}\approx 0.68$, $\Omega_{m0}\approx0.32$ and $\Omega_{r0}\approx 10^{-5}$ at face value for the numerical integrations (in line with local direct measurements as those reported in reference \cite{Riess:2021jrx}).

\begin{figure*} 
    \begin{subfigure}{0.4\textwidth}
    \begin{adjustwidth}{0cm}{}
    \includegraphics[scale=0.71]{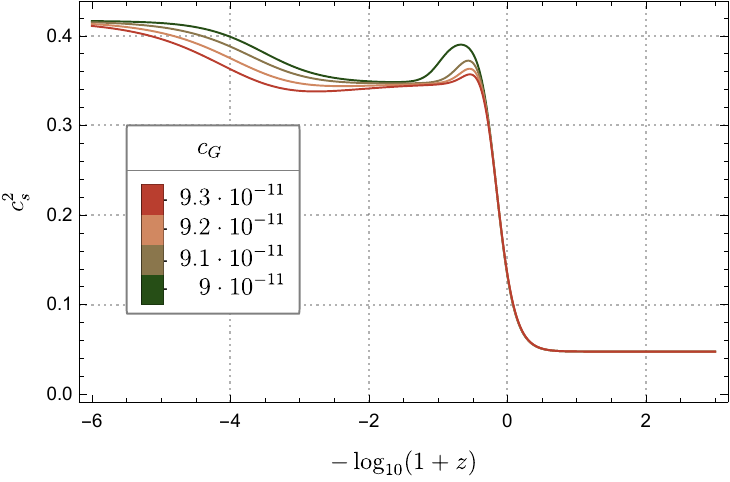}
    \end{adjustwidth}
    \end{subfigure}
    \hfill
    \begin{subfigure}{0.4\textwidth}
    \begin{adjustwidth}{-1.7cm}{}
    \includegraphics[scale=0.72]{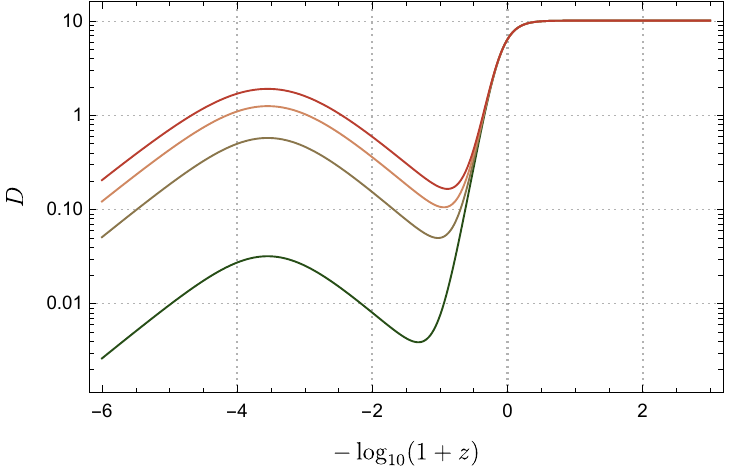}
    \end{adjustwidth}
    \end{subfigure}   
\caption{Numerical evolution for $\eta=0.4$ and  $\alpha=0.6$ in expressions (\ref{eq:cteETA}) and (\ref{eq:K}), respectively. In addition, the coupling constant $c_K$ equals minus unity, whereas $c_G$ ranges from $9\cdot10^{-11}$ to $9.3\cdot10^{-11}$. Left panel represents the speed of sound squared, $c_s^2$, as defined in equation (\ref{eq:gradstab}). Right figure shows  the evolution of the ghost parameter $D$ defined in equation (\ref{eq:scalarstab}). As can be seen, linear order perturbations (where scalar perturbations are decoupled form GWs) are always stable even though the scalar field displays a phantom behaviour at present epoch. \label{fig:mod_1 2}}
\end{figure*}

\begin{figure}
    \centering
    \includegraphics[width=\columnwidth]{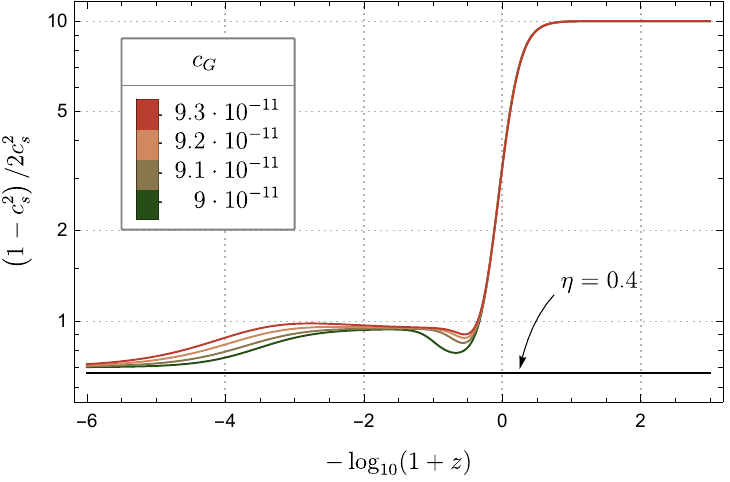}
    \caption{Numerical evolution for $\eta=0.4$ and  $\alpha=0.6$ in expressions (\ref{eq:cteETA}) and (\ref{eq:K}). Taking $c_K=-1$, the curves represent  the ratio $(1-c_s^2)/2c_s^2$ from equation (\ref{eq:noGWsGrad}) for different values of the coupling constant $c_G$. The horizontal line depicts the upper-bound for this ratio found in expression (\ref{eq:noGWsGrad}). Clearly the model never satisfies this bound for the domain of the numerical integration and, therefore, the interaction with GWs always introduce (at least) a gradient instability in the scalar sector.}
    \label{fig:pGWs_mod1}
\end{figure}

\begin{figure*}
    \begin{subfigure}{0.4\textwidth}
    \begin{adjustwidth}{0cm}{}
    \includegraphics[scale=0.7]{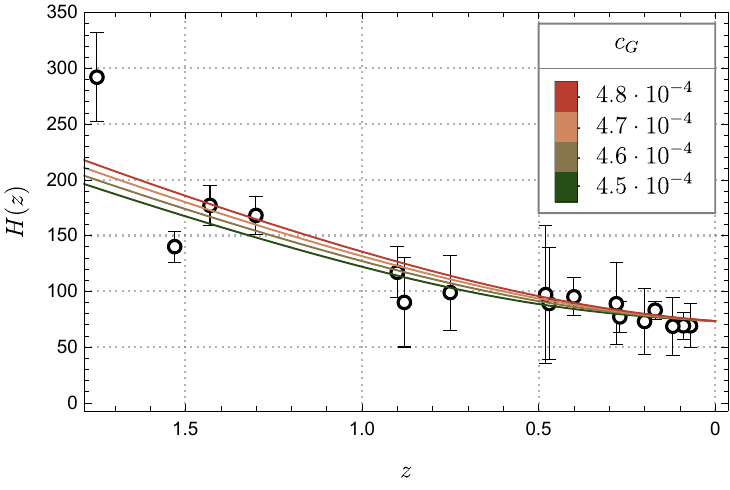}
    \end{adjustwidth}
    \end{subfigure}
\hfill
    \begin{subfigure}{0.4\textwidth}
    \begin{adjustwidth}{-1.7cm}{}
    \includegraphics[scale=0.7]{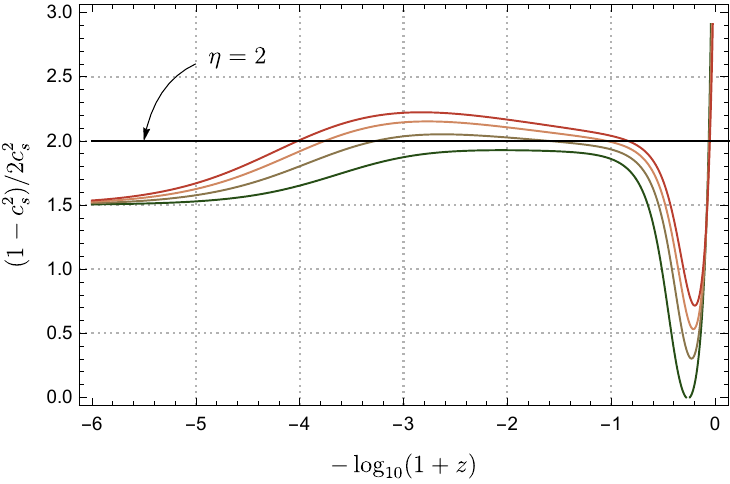}
    \end{adjustwidth}
    \end{subfigure}
\caption{Numerical evolution for $\eta=2$ and  $\alpha=2$ in expressions (\ref{eq:cteETA}) and (\ref{eq:K}), respectively. In addition, the coupling constant $c_K$ equals minus unity, whereas $c_G$ ranges from $4.5\cdot10^{-11}$ to $4.8\cdot10^{-11}$. Left panel shows the evolution of the Hubble rate at low redshift compared to direct measurements collected from references \cite{2014RAA,Simon:2004tf,Ratsimbazafy:2017vga,2010JCAP,Borghi:2021rft} (see also Table II in reference \cite{Alonso-Lopez:2023hkx}). Right panel represents the ratio $(1-c_s^2)/2c_s^2$ from equation (\ref{eq:noGWsGrad}). The horizontal line there depicts the upper-bound this ratio should satisfy to avoid a GWs-induced gradient instability; see inequality (\ref{eq:noGWsGrad}). \label{fig:mod_2}}
\end{figure*}

Our numerical analysis has shown two main tendencies for these models. First, we found that the ghost-free and gradient-free conditions discussed in section \ref{sec:linearPerturbs} (where scalar perturbations decouple from GWs at linear order) are not fulfilled for all the parameter-space, but only in some regions. Second, for the models that satisfy the ghost-free and gradient-free at linear order, we have found that the condition (\ref{eq:noGWsGrad}) is always violated at some point during the (past) evolution of the system. 
We summarise these findings with the two proxy KGB candidates shown in figures \ref{fig:mod_1} to \ref{fig:mod_2}. In the former, we consider $\eta=0.4$, $\alpha=0.6$ and $c_G$ ranging form $9\cdot10^{-11}$ to $9.3\cdot10^{-11}$. As can be appreciate in figure \ref{fig:mod_1}, this simple KGB model has a compelling behaviour on the background level. Depending on the value of the coupling constant $c_G$, the fractional contribution of the scalar field to the total energy of the universe picks at matter-radiation equality, rendering this model as a potentially interesting Early Dark Energy (EDE) candidate (see a review on EDE in, for instance, reference \cite{Poulin:2023lkg}). Moreover, the resulting evolution of the Hubble rate at low redshift can be, in principle, made compatible with the current data available from direct measurements\footnote{We recall that  we have not fitted the parameters of the model at hands with observational data. Contrary, the previous statement comes from just comparing the resulting $H(z)$ with what to be expected taking into account the current data coming from direct measurements collected from references \cite{2014RAA,Simon:2004tf,Ratsimbazafy:2017vga,2010JCAP,Borghi:2021rft}.}. At the level of linear scalar perturbations, the ghost-free and gradient-free conditions are always satisfied, even though the model at hands shows a slight phantom behaviour at present time; see figure \ref{fig:mod_1 2}. Nevertheless, as can be seen in figure \ref{fig:pGWs_mod1}, the condition (\ref{eq:noGWsGrad}) is never fulfilled for the domain of the numerical integration (that is from $z_i=10^6$ to $z_f=-0.999$). Since this bound is always violated, the signs of $Z^{11}$ and $Z^{22}$ will alternate between positive and negative values if $|\,\beta\,|>1$. In that case, conditions (\ref{cond:gradZ11}) and (\ref{cond:gradZ22}) would not always be satisfied during the period of a single oscillation of the GWs. As a result, the interaction between DE fluctuations and GWs will induce (at least) a gradient instability for certain (periodic) values of the argument $\omega(t-z)$ if the amount of braiding is non-negligible. Hence, this model cannot provide a working example for the mechanism we discussed in section \ref{sec:noGrad}.

A second example is shown in figure \ref{fig:mod_2}. This corresponds to $\eta=2$,  $\alpha=2$ and $c_K=-1$ in expressions (\ref{eq:cteETA}) and (\ref{eq:K}). In this case, the coupling constant $c_G$ takes values from $4.5\cdot10^{-4}$ to $4.8\cdot10^{-4}$. The background evolution of this proxy model is still compatible with low redshift measurements, but shows some tensions with data above $z\approx 1$ for some values of $c_G$; see left panel in figure \ref{fig:mod_2}.  The right panel shows the fulfilment of the condition (\ref{eq:noGWsGrad}). This condition is always satisfied in the asymptotic past of the model (within the limits of the integration domain). Unfortunately, it is consistently violated when the scalar field becomes dominant, if not even before that moment. As a result, interaction between DE fluctuations and GWs will induce (at least) a gradient instability if the amount of braiding is non-negligible.

The models represented in figures \ref{fig:mod_1} to \ref{fig:mod_2} illustrate the general behaviour we found for the functions (\ref{eq:cteETA}) and (\ref{eq:K}). Even though there exist regions in the parameter-space for which $D>0$ and $c_s^2\geq0$ always, the condition (\ref{eq:noGWsGrad}) is, in the best case scenario, fulfilled only for a finite time during the (past) evolution of the model. Moreover, this condition is systematically violated (at the latest) when the scalar field becomes dominant. Hence, a GWs-induced instability, like that described in reference \cite{Creminelli3} if the amount of braiding is non-negligible, seems to be inevitable for the models at hands.

However, it is important to emphasize that the numerical screening of the $(\eta,\alpha)$-plane we have performed is not an analytic proof of the non-viability of the KGB example given by the functions (\ref{eq:cteETA}) and (\ref{eq:K}). Moreover, a different choice for the $K$ function may also affect the results. Nevertheless, given the naturalness of power-law functions in physics and the lack of a positive conclusion in our numerical analysis, we suspect that finding a KGB model that could conduct the mechanism we discussed in section \ref{sec:noGrad} for avoiding a GWs-induced instability \cite{Creminelli3}  would be rather difficult, if not impossible.

\section{What is left?\label{sec:conclusions}}

The braiding term $G\Box\phi$ in the scalar field action (\ref{eq:actionKGB}) has been shown to interact with GWs in such a way that it produces a ghost and/or a gradient instability in the scalar perturbations if the amount of braiding present at cosmological scales is non-negligible \cite{Creminelli3}. Motivated by the fact that the stability analysis in reference \cite{Creminelli3} was mainly performed for the special case of cubic Galileon, i.e. $G\propto X$, in this paper we have proposed a theoretical mechanism to avoid these GW-induced instabilities for KGB models beyond cubic Galileon. This mechanism relies on having a large and positive $\eta$ (quantity that measure deviations w.r.t. a linear function $G$) for expanding the stability region to $|\,\beta\,|\gg1$ at low values of the speed of sound, as shown in figure \ref{fig:stabGWs}. It should be mentioned, however, that the parameter $\beta$ is not only proportional to the amount of braiding in the theory (through $\bar{m}_1^3$ or, equivalently, $\alpha_{\rm B}$) but to the strain amplitude of the GWs as well, see equation (\ref{beta}). If observational data of GWs events with higher and higher strain amplitudes are considered, then the resulting $\bar{m}_1^3$ (or $\alpha_{\rm B}$) will still be constrained to be negligible, even though our mechanism could extend the stability region to larger values of $\beta$. In that scenario, the conclusion in reference \cite{Creminelli3} about the non-viability of the braiding at cosmological scales will be only marginally affected by the discussion we have proposed here. 
Nevertheless, before finally concluding that the KGB theory should not have any sizeable effects, the fate of these instabilities, once originated, should be addressed. This lays beyond the scope of the present work.

We have also explored whether some reasonable KGB candidate could conducted the theoretical mechanism proposed  here for evading GWs-induced instabilities.
Unfortunately, using numerical simulations with a power-law prescription for the functions $K$ and $G$ we have not been able to find a working example for this effect. As we commented in section \ref{sec:noExample}, the power-law choice for $G$ provides a physically relevant case for our mechanism to work. Therefore, we consider the lack of a working example with power-law functions as a solid hint towards the non-viability of a cosmological braiding like the one present in the shift-symmetric KGB theories. 

Conversely, the $k$-essence part in the KGB action (\ref{eq:actionKGB}) trivially evades the constraints coming from references \cite{Creminelli3,Creminelli1,Creminelli2}. Moreover, this scalar field theory can feature  ghost-free and gradient-free scalar perturbations if the NEC is satisfied for the DE sector. As we have discussed, however, the ghost-free and gradient-free conditions for $k$-essence are anti-correlated when the NEC is violated. Therefore, if we were to describe a DE component that violates the NEC (a possibility that is observationally viable \cite{Risaliti:2018reu}) with a single scalar field within the $k$-essence theory, then we will inevitably face a ghost or a gradient instability. Whether the instability will be ghost- or gradient-like depends, ultimately, on the function $K$. It is also important to note that $k$-essence cannot produce a phantom crossing \cite{Vikman:2004dc,Caldwell:2005ai}. Hence, the scalar field will always be phantom or non-phantom, with the stability issues this may entail in the former case. This points towards the impossibility of solving simultaneously the $H_0$ and $S_8$ tensions with late-time modifications to $\Lambda$CDM \textit{\`a la} $k$-essence should the $S_8$ tension be still present in future observations, as this resolution would demand a phantom crossing \cite{Heisenberg:2022lob,Heisenberg:2022gqk}.  See also previous works on the $H_0$ tension in the $k$-essence theory in, for instance, references \cite{Banerjee:2020xcn,Lee:2022cyh}.
With this potential limitation in mind, an interesting extension of the $k$-essence scenario is that when more than one scalar field is present \cite{Malik:2008im,Wands:2007bd}. In that case, the effective DE fluid can cross the phantom divide \cite{Hu:2004kh} even though each scalar field, namely $\phi_i$, is restricted to the phantom or non-phantom regime in the same way as described in section \ref{sec:kessence}.

It should be noted that the discussion presented in this work will be only partially affected if we allow the shift-symmetry to be (weakly) broken at some regimes. This would introduce an explicit dependence on the scalar field $\phi$ in the functions $K$ and $G$. The destabilising interaction between GWs and the braiding term will still occur even for $G=G(\phi,X)$ \cite{Creminelli3}. However, it would be interesting to explore whether the mechanism proposed in section \ref{sec:noGrad} will have a better performance in that case.
If not, then,  the braiding term should not have any sizeable effect  and the surviving theory will be effectively that of $k$-essence. 
For this non-shift-symmetric $k$-essence theory, no additional terms will appear in the formulas for the scalar field energy density and pressure, or in the equations for $c_s^2$ and $D$; see the corresponding expressions in, for instance, reference \cite{KGB1}. Thus, our discussion about the linear stability of the theory shall remain the same as for the shift-symmetric $k$-essence case. In addition to the previous considerations, the $k$-essence scalar field cannot produce a phantom crossing even if $K=K(\phi,X)$; see reference \cite{Vikman:2004dc,Caldwell:2005ai}. Therefore, the conclusions presented in section \ref{sec:kessence} shall remain unchanged.

Along this line of thought, another possible extension of the $k$-essence theory  is that when a non-minimal coupling to gravity is allowed for the scalar field.  Assuming that the external matter field are minimally coupled to the metric $g_{\mu\nu}$, the gravitational part of the total action would read
\begin{align}\label{eq:action_left}
		S_g=\int d^4 x\sqrt{-g}\left[f(\phi) R+ K(\phi,X) \right],
\end{align}
where the non-minimal coupling function $f$ can only depend on the scalar field, but not on its kinetic term $X$. This is due to strong restrictions on the speed of propagation of GWs; see reference \cite{Kase:2018aps}. Please note that this scalar field theory would also trivially evade the constraints from \cite{Creminelli3}. Nevertheless, the non-minimal coupling may introduce a fifth-force at local scales. Therefore, some mechanism should be invoked to screen this force on astrophysical scales; see, for instance, references \cite{Brax:2014yla,Crisostomi:2019yfo,DeFelice:2011th} and references therein. This limitation may also be present in extensions of $k$-essence to modified theories of gravity like those discussed in, e.g., references \cite{Noh:2001ia,Hwang:2002fp,Hwang:2005hb,Chervon:2018ths,Nojiri:2019dqc,Odintsov:2020qyw}.

Furthermore, within the spirit of analysing fundamental symmetries in order to address DE, we should also briefly mention the possibility of considering a scalar field with an action that is invariant just under transverse diffeomorphisms
\cite{Alonso-Lopez:2023hkx,Alvarez:2009ga,Maroto:2023toq,Jaramillo-Garrido:2023cor}. In this case, one can describe phenomenology of interest for the dark sector even with a shift-symmetric field with a canonical kinetic term \cite{Alonso-Lopez:2023hkx,Jaramillo-Garrido:2023cor}. On the other hand, one can consider more general theories invariant only under transverse diffeomorphisms \cite{Bello-Morales:2023btf,Maroto:2024mkx}, being the most prominent example that of Unimodular Gravity \cite{Einstein} (see also \cite{Henneaux:1989zc,PhysRevD.40.1048,Alvarez:2005iy,Carballo-Rubio:2022ofy, Alvarez:2023utn,Salvio:2024vfl}). Finally, as it is well-known, more general alternative theories of gravity goes beyond the consideration of a scalar field to describe the dark sector \cite{CANTATA:2021ktz}. 

\change{Finally, we shall recall again that we have discussed the viability of the KGB theory solely from the point of view of the stability of scalar perturbations. In doing so, we have focus on theoretical considerations only. Should any of the models studied had fulfilled our stability criteria, then the analysis of background expansion by means of the evolution of the partial energy densities and the Hubble rate (like shown in figures \ref{fig:mod_1} and \ref{fig:mod_2}) would not be sufficient to claim for the viability of the DE-candidate. In such case, a thorough analysis on the observational impact using more advanced techniques and specialised software like CLASS should be carried on; see, for instance, references \cite{Zumalacarregui:2016pph,Bellini:2019syt}. We leave for future research the stability of more involving KGB models and their compatibility with observational data. }


\bigskip
\noindent\textbf{CRediT authorship contribution statement}
\bigskip

\textbf{Teodor Borislavov Vasilev:} Conceptualization, Investigation, Formal analysis, Software, Validation, Visualization, Writing – original draft, Writing – review \& editing. \textbf{Mariam Bouhmadi-López:} Conceptualization, Supervision, Validation, Writing – review \& editing. \textbf{Prado Martín-Moruno:} Conceptualization, Supervision, Validation, Writing – review \& editing.

\bigskip
\noindent\textbf{Declaration of competing interest}
\bigskip

The authors declare that they have no known competing financial interests or personal relationships that could have appeared to
influence the work reported in this paper.

\bigskip
\noindent\textbf{Data availability}
\bigskip

No data was used for the research described in the article.

\bigskip
\noindent\textbf{Acknowledgments}
\bigskip

    The research of T.B.V. and P.M.M. is supported by the project PID2022-138263NB-I00 funded by MICIN/AEI/10.13039/501100011033  and by FEDER, UE. 
    T.B.V. also acknowledge financial support from Universidad Complutense de Madrid and Banco de Santander through Grant No. CT63/19-CT64/19. 
    The work of M.B.L. is supported by the Basque Foundation of Science Ikerbasque. Her work has been also financed by the Spanish projects PID2020-114035GB-100 (MINECO/AEI/FEDER, UE) \change{and PID2023-149016NB-I00
    (MINECO/AEI/FEDER, UE)}. She would like to acknowledge the financial support from the Basque government Grant No. IT1628-22 (Spain).


\appendix

\section{Stability of linear scalar perturbation \label{app:FKGB}}

    In this appendix we discuss the possibility of satisfying simultaneously the ghost and gradient-free conditions for linear scalar perturbations given by the inequalities (\ref{eq:ghoststabFKGB}) and (\ref{eq:gradstabFKGB}), respectively.
    In order to make our notation more compact, it will be useful to introduce the two following reference scales
	\begin{align}
		\Omega^\star&\coloneqq\frac{(3+\Omega_r-3\Omega_\phi)\Omega_J^B}{3(2-\Omega_J^B)},\\
		\bar{\Omega}&\coloneqq\frac{2+\eta}{3+\eta}(1+w_\phi)\Omega_\phi-\frac{\Omega_J^B(2-\Omega_J^B)}{6(3+\eta)},
	\end{align}
	where $\Omega^\star$ is well-defined only for $\Omega_J^B\neq2$. The case of $\Omega_J^B=2$ can be directly addressed using the expressions (\ref{eq:ghoststabFKGB}) and (\ref{eq:gradstabFKGB}). We consider here the parameter $\eta$ to be positive since that is the case proposed in section \ref{sec:noGrad} for avoiding a GWs-induced gradient instability. Also note that the quantity $\Omega_r+3-3\Omega_\phi$ is always positive\footnote{This is because $\Omega_r+3-3\Omega_\phi=4\Omega_r+3\Omega_m$, which is clearly positive since the energy densities of radiation and matter are non-negative.}. Hence, $\Omega^\star$ is positive if $\Omega_J^B\in(0,2)$, whereas it is negative for $\Omega_J^B<0$ and $\Omega_J^B>2$.
	
	Combining the ghost and the gradient-free conditions lead to the following scenarios in which both conditions are satisfied for $\eta$ non-negative:
    \paragraph{Scenario 1} ($\Omega_J^B<0,\, \Omega_J>\max\{(1+w_\phi)\Omega_\phi,\,\Omega^\star,\,\bar{\Omega}\}$): Both the numerator and the denominator in equation (\ref{eq:ghoststabFKGB}) are positive. The reference scale $\Omega^\star$ is negative. Phantom behaviour is allowed.
    \paragraph{Scenario 2} ($0<\Omega_J^B<2,\, \max\{(1+w_\phi)\Omega_\phi,\,\bar{\Omega}\}<\Omega_J<\Omega^\star$): The numerator and the denominator in equation (\ref{eq:ghoststabFKGB}) are positive. The reference scale $\Omega^\star$ is also positive. Clearly,  $\max\{(1+w_\phi)\Omega_\phi,\,\bar{\Omega}\}<\Omega^\star$ should hold for consistency. Phantom behaviour is allowed. 
    \paragraph{Scenario 3} ($0<\Omega_J^B<2,\,\max\{\Omega^\star,\bar{\Omega}\}<\Omega_J<(1+w_\phi)\Omega_\phi$): The numerator and the denominator in equation (\ref{eq:ghoststabFKGB}) are negative. The reference scale $\Omega^\star$ is positive and, therefore, so it is $\Omega_J$.  Hence, from the upper bound for $\Omega_J$ it follows that the scalar field cannot be phantom.
    \paragraph{Scenario 4} ($\Omega_J^B=2, \Omega_J>\max\{(1+w_\phi)\Omega_\phi,\,\bar{\Omega}\}$): The numerator and the denominator in equation (\ref{eq:ghoststabFKGB}) are positive. Note that $\bar{\Omega}>(1+w_\phi)\Omega_\phi$ if the scalar field is phantom-like and, therefore, $\Omega_J$ should be greater that $\bar{\Omega}$. In the non-phantom regime, the lower bound for $\Omega_J$ is given by  $(1+w_\phi)\Omega_\phi$. 
    \paragraph{Scenario 5} ($\Omega_J^B>2,\, \Omega_J>\max\{(1+w_\phi)\Omega_\phi,\,\Omega^\star,\,\bar{\Omega}\}$): The numerator and the denominator in equation (\ref{eq:ghoststabFKGB}) are positive. The reference scale $\Omega^\star$ is negative. Phantom behaviour is allowed.

    \bigskip
    Even though the physical intuition behind each of these scenarios is not so transparent as for the marginal models studied in sections \ref{sec:kessence} and \ref{sec:PB}, their very existence is a solid proof that it is possible to have a ghost-free and gradient-free scalar perturbations at linear level (recall we consider here only the case of $\eta$ non-negative) in the shift-symmetric KGB theory even when $w_\phi<-1$.
\section{Quick guide to the EFT of dark energy\label{app:EFT}}

    The framework of EFT was first applied to DE in reference \cite{Creminelli:2008wc} (see also \cite{Cheung:2007st,Weinberg:2008hq} for applications to inflationary models) and further developed in references \cite{Gubitosi:2012hu,Bloomfield:2012ff,Gleyzes:2013ooa,Piazza:2013coa} (see also reference \cite{Frusciante:2019xia} for a review). 
    This approach considers the most general form for the gravitational action (including cosmological perturbations up-to an arbitrary order) built-up only on symmetry arguments. Originally, this framework was based on two assumptions \cite{Gubitosi:2012hu}: $i)$ the validity of the weak equivalence principle and the existence of a Jordan metric $g_{\mu\nu}$ universally coupled to matter fields; $ii)$ the existence of a unitary gauge compatible with the residual symmetries of unbroken spatial diffeomorphisms.

    The last of these assumptions advocates for the presence of a scalar field $\phi$ in the DE sector. This is so because the presence of a homogeneous scalar field $\phi(t)$ in a FLRW background defines a preferred time slicing of space-time. These are the hypersurfaces with $\phi$ constant and, therefore, $\delta\phi\equiv0$ on them. So, for this choice (unitary gauge) only metric degrees of freedom are explicitly displayed in the action. The scalar field perturbation can be reintroduced explicitly in the theory via the St\"ueckelberg trick. That is, by performing infinitesimal time diffeomorphism  $t\to t+\pi(t,x)$ being $\pi$ the scalar field fluctuations. However, in this approach $\pi$ does not represent the original scalar field $\phi$ in the DE sector, but the perturbations encoding the scalar degree of freedom in the theory.
	
    In order to construct the most general expression for the action satisfying the previous ans\"atze, note that hypersurfaces with constant $\phi$ can be defined as those orthogonal to the unit four-vector
	\begin{align}\label{def:slicing}
		n_\mu\coloneqq-\epsilon\frac{\partial_\mu\phi}{\sqrt{2X}},
	\end{align}
    provided that $\partial_\mu\phi$ is time-like, i.e. $X>0$,  where $\epsilon\coloneqq \textup{sgn}(\dot\phi)=\pm1$ is to ensure the four-vector is future-oriented. (Here we are not going to consider oscillating solutions where the background field velocity, $\dot\phi$, crossed zero since this case may be problematic in standard perturbation theory \cite{Bellini:2014fua}.) For a homogeneous scalar field, that is $\phi=\phi(t)$, this four-vector reduces to
	\begin{align}
		n_\mu=-\frac{\delta_\mu^0}{\sqrt{-g^{00}}}.
	\end{align}
    This slicing induces the spatial metric
	\begin{align}
		h_{\mu\nu}=g_{\mu\nu}+n_\mu n_\nu,
	\end{align}
	where $n_\mu n^\mu=-1$ and $n^\mu h_{\mu\nu}=0$. 
	
	Note that invariance under time-translations is spontaneously broken (in the sense discussed in reference \cite{Piazza:2013coa}) since the scalar field signals out a preferred time. The terms allowed in the EFT action are, therefore, those invariant under the residual symmetries of the unbroken spatial diffeomorphisms, such as the contravariant time-time component of the Jordan metric $g^{00}$. 	
	The extrinsic curvature of constant time hypersurfaces is also allowed to appear. This is defined as \cite{Wald:1984rg}
	\begin{align}
		\K_{\mu\nu}\coloneqq h_\mu^{\ \alpha}\nabla_\alpha n_\nu,
	\end{align}
    and represents the spatial projection of the covariant derivative of $n_\nu$. (Note that the extrinsic curvature is sometimes defined with a different sign convention, see, for instance, references \cite{Burrage:2010cu,Capozziello:2011et}.) In addition, its trace reads
	\begin{align}
		\K=\nabla_\mu n^\mu.
	\end{align}
    It will be useful for latter calculations to note that
    \begin{align}\label{eq:int lK}
	&\int\textup{d}^4x\sqrt{-g} \, l(t) \K = \int\textup{d}^4x\sqrt{-g}\, l(t) \nabla_\mu n^\mu \nonumber\\
        &= - \int \textup{d}^4x\sqrt{-g}\, n^\mu  \partial_\mu l = - \int \textup{d}^4x\sqrt{-g}  \sqrt{-g^{00}}\, \dot l\, ,
    \end{align}
    up-to boundary terms and for any function $l$ on time. 
    
    The Ricci scalar $R$, any curvature invariants, and contractions of tensors with $g_{\mu\nu}$, $n_\mu$ and the covariant derivative $\nabla_\mu$ are also invariant under spatial diffeomorphisms.
    Therefore, the most general EFT action can be expressed as \cite{Cheung:2007st} 
    \begin{align}
		S_g=\int\textup{d}^4x\sqrt{-g}\mathcal{L}\left(R_{\mu\nu\rho\sigma},\K_{\mu\nu},g^{00},t\right),
    \end{align}
    where time is also allowed to appear explicitly. For addressing the cosmological perturbation, however, it is more convenient to re-write the above action into the part contributing to the background level and that enclosing the perturbations around a flat FLRW metric (at any order). This is \cite{Gubitosi:2012hu} (sea also \cite{Cheung:2007st,Gleyzes:2013ooa}) 
    \begin{align}\label{actionEFTbackground}
		S_g=\int\textup{d}^4x\sqrt{-g}\left[\frac{M^2_*}{2}f(t)R-\Lambda(t)-c(t)g^{00}\right]+S^{(2)}_{DE},
    \end{align}
    where $f$, $\Lambda$ and $c$ are functions of the cosmic time $t$. Moreover, the latter two background quantities read
    \begin{align}
		c &  =  \frac{M^2_*}{2} (  - \ddot f  +  H \dot f ) + \frac12 (\rho_{\textup{DE}}+p_{\textup{DE}}), \label{eq:c(t)}\\
		\Lambda &  = \frac{M^2_*}{2}(   \ddot f  + 5 H \dot f ) + \frac12 (\rho_{\textup{DE}}- p_{\textup{DE}}), \label{eq:Lambda(t)}
    \end{align}
    being $p_{\textup{DE}}$ and $\rho_{\textup{DE}}$ the pressure and energy densities of the DE fluid. In addition, $M^2_*$ is the Planck mass. Hereon we will take $M^2_*=1$ since we have adopted the geometric unit system in this paper.
    Conversely, $S^{(2)}_{DE}$ contains all terms that start at quadratic order in perturbations and, therefore, they do not affect the background. (Note, however, that the background quantities do affect all perturbation levels.) This part can be expressed as follows \cite{Gubitosi:2012hu}
    \begin{align}\label{actionEFT}
		S^{(2)}_{DE}=\,\frac12\int\textup{d}^4x&\sqrt{-g}\bigg[M_2^4 (\delta g^{00})^2 
		- \bar m_1^3\,  \delta g^{00} \delta \K - \bar M_2^2\,  \delta \K^2 \nonumber\\
        &- \bar M_3^2\,  \delta \K_{\mu}^{\ \nu} \delta \K_{\ \nu}^\mu + m_2^2 h^{\mu \nu} \partial_\mu g^{00} \partial_\nu g^{00}\nonumber\\
		&+\lambda_1 \delta R^2 + \lambda_2 \delta R_{\mu \nu} \delta R^{\mu \nu} + \mu_1^2 \delta g^{00} \delta R\nonumber\\
        &+\gamma_1  C^{\mu \nu \rho \sigma} C_{\mu \nu \rho \sigma} + \gamma_2  \epsilon^{\mu \nu \rho \sigma} C_{\mu \nu}^{\ \ \ \kappa \lambda} C_{\rho \sigma  \kappa \lambda}\nonumber \\
		&+ \frac{M_3^4}{3} (\delta g^{00})^3 - \bar m_2^3\,  (\delta g^{00})^2  \delta \K  + \dots \bigg],
    \end{align}
    where $M^4_i,\,\bar{M}^2_i,\,m^2_i,\,\bar{m}^3_i$ and $\mu_i$ are mass parameters whereas $\lambda_i$ and $\gamma_i$ are dimensionless quantities. Generally, these parameters are allowed to depend on the time coordinate $t$. In addition, $\delta g^{00}\coloneqq g^{00}+1$ is the perturbation of the metric, $\delta \K^{\mu}_{\,\nu}$ is the perturbation of the extrinsic curvature, and $\delta \K$ the perturbation of its trace. Likewise,  $\delta R_{\mu\nu}$ is the perturbation of the Ricci tensor and $\delta R$ the perturbation of its trace, whereas $C_{\mu\nu\rho\sigma}$ represents the Weyl tensor. The action (\ref{actionEFTbackground}) describes virtually all DE models encompassing a single scalar degree of freedom.

    For the case of the KGB theories given by action (\ref{eq:actionKGB}), the background quantities in (\ref{actionEFTbackground}) read \cite{Gubitosi:2012hu}
    \begin{align}
		f=&\,1,\\
		c=&\,\frac12(\rho_{\phi}+p_{\phi}),\label{c_EFT}\\
		\Lambda=&\,\frac12(\rho_{\phi}-p_{\phi})\label{Lambda_EFT},
    \end{align}
    with $\rho_\phi$ and $p_\phi$ given by equations (\ref{rho_phi}) and (\ref{p_phi}), respectively. Whereas the action (\ref{actionEFT}) for the perturbations reduces to \cite{Gubitosi:2012hu}
    \begin{align}\label{actionEFT_KGB}
		S^{(2)}_{DE}=\,\frac12\int&\textup{d}^4x\sqrt{-g}\bigg[M_2^4 (\delta g^{00})^2 
		- \bar m_1^3\,  \delta g^{00} \delta \K\nonumber\\
        &+ \frac{1}{3}M_3^4 (\delta g^{00})^3 - \bar m_2^3\,  (\delta g^{00})^2  \delta \K  + \dots \bigg] .
    \end{align}
    That is, only the mass parameters $M_{n}^4$ and $\bar{m}_{n-1}^3$ (for $n\geq2$) are present for the KGB theory (see also, for instance, reference \cite{Frusciante:2019xia}). Recall that these parameter are, in general, functions on the time coordinate $t$. Moreover, note that  $M_2^4$ is connected with the \textit{kineticity} term $\alpha_K$, whereas $\bar{m}_1^3$ is equal to the braiding term $\alpha_B$ up-to sign conventions; see the discussion in \ref{app:EFT FKGB} (see also table 2 in reference \cite{Bellini:2014fua}). It is also important to highlight that the operator introducing the kinetic mixing with gravity at leading order is $\bar{m}_{1}^3$. This can be understood in an intuitive way by noting that the perturbation $\delta g^{00}$ introduces a term proportional to $\dot\pi$ at linear order after the St\"uckelberg procedure, see equation (\ref{eq:delta g00}). The extrinsic curvature, on the other hand, contains time derivative of the spatial metric, $h_{ij}$, when time diffeomorphism is fully restored \cite{Wald:1984rg} (see also, for instance, references \cite{Gubitosi:2012hu,Gleyzes:2014rba,Cai:2016thi}). Schematically, the product $\delta g^{00}\delta\K$, therefore, introduces a term proportional to $\dot\pi\dot h$ (at second order) into the Lagrangian. In the Newtonian gauge (\ref{eq:NewtonianG}), this contribution is proportional to $\dot\pi\dot \Psi$; see reference \cite{Gubitosi:2012hu}. For the same reason, the operators $\bar{m}_{n-1}^3$ (for $n\geq2$) will also be responsible for introducing the kinetic mixing between  scalar perturbations and the GWs at higher  orders.

    In the following part of the appendix, we compute the expression of these mass parameters in terms of the KGB functions $K$ and $G$, and their derivatives. These calculations will be done without invoking a shift-symmetry and, therefore, the resulting expression are valid for the most general KGB scenario.


    \subsection{Mass parameters for \textit{k}-essence \label{sec:appK}}
    
    As a warm-up, we consider first the $k$-essence theory. In the most general scenario, the $k$-essence function depends on both the scalar field and its kinetic term; i.e. $K=K(\phi,X)$. (We remind the reader that we will not invoke a shift-symmetry here; thus, the results obtained in this appendix are valid for the most general case.) When perturbing this function, one may naively consider that perturbations around the background values of both $\phi$ and $X$ should be taken into account. However, in the unitary gauge (defined by $\delta\phi\equiv0$) only the perturbations of the kinetic term remains. 
    Let the perturbed kinetic term be defined as
    \begin{align}
        \X\coloneqq X+\delta X,
    \end{align}
    where $X$ is the background value and $\delta X$ the perturbation. In the unitary gauge, the perturbed kinetic term $\X$ is related to the perturbed metric through
    \begin{align}\label{def:Xpertub2}
		\X=&-\frac12\dot{\phi}^2g^{00}=-Xg^{00},
    \end{align}
    being $g^{00}$ the perturbed time-time component of the metric. Hence, 
    \begin{align}\label{def:Xpertub3}
        \delta X=&-X\delta g^{00},
    \end{align}
    in the unitary gauge. Consequently, the perturbed $k$-essence function becomes
    \begin{align}
        K(\phi,\X) \xrightarrow{\text{unitary gauge}} K(t,\,g^{00}),
    \end{align}
    where we have assumed that $\phi$ and its kinetic term are functions of time only at the FLRW background. Note that the scalar field does not appear explicitly in the unitary gauge: it is ``eaten'' in the metric degrees of freedom \cite{Frusciante:2019xia,Gubitosi:2012hu}. However, in order to obtain the expressions for the mass parameters in the EFT language in terms of $K(\phi,X)$ and its derivatives, it will be useful to maintain the usual covariant notation for the function $K$ and its arguments instead of $K(t,\,g^{00})$ when there is no risk for confusion. Please note that this is an abuse of notation since neither $\phi$ nor $X$ can appear explicitly in this gauge. They are simply functions on the time coordinate.

    In this notation, the $k$-essence function can be expanded around its background value as 
    \begin{align}\label{eq:EFTkessence}
         K(\phi,\X)= K(\phi,X)+\sum_{n=1}^\infty\frac{1}{n!}\left.\diffp[n]{K(\phi,\X)}{\X}\right|_{\delta X=0}(\delta X)^n,
    \end{align}
    where subscript ``$\delta X=0$" denotes evaluation of the corresponding quantity on the background level. For the sake of the compactness of the notation, we will avoid writing explicitly the arguments of the function $K$ and its derivatives appearing in the r.h.s. of the expansion, where it should be kept in mind that these expressions are always evaluated on the background. We will also omit the subscript ``$\delta X=0$" in the following and express the coefficient in the expansion simply as $\partial^n K/\partial X^n$ where there is no risk for confusion.
    Accordingly, the above expansion can be rewritten as
    \begin{align}\label{eq:EFTkessence3}
		K(\phi,\X)=K-XK_{X}-XK_{X}g^{00}+\sum_{n=2}^\infty\frac{(-X)^n}{n!}\diffp[n]{K}{X}(\delta g^{00})^n,
    \end{align}
    where we have substituted $\delta X=\X-X$ in the first order in the expansion, and used equations (\ref{def:Xpertub2}) and (\ref{def:Xpertub3}) to explicitly introduce $g^{00}$ and $\delta g^{00}$, respectively. Comparing the above expression with equations (\ref{actionEFTbackground}) and (\ref{actionEFT}), it is straightforward to readout the contribution of the $k$-essence function to the EFT parameters. These are\footnote{\label{fn:Gub}Please, note that the authors in reference \cite{Gubitosi:2012hu} used a different convention for the definition of the kinetic term. This is $X\coloneqq g^{\mu\nu}\nabla_\mu \phi\,\nabla_\nu\phi$ \cite{Gubitosi:2012hu}, in contrast with our definition for $X\coloneqq-\frac12 g^{\mu\nu}\nabla_\mu \phi\,\nabla_\nu\phi$, where the convention used for the signature of the metric tensor is in both cases the same. Also note that it is always possible to redefine the scalar field in a way that $\phi(t)=\sqrt{2}t$ and, therefore, the kinetic term would simply read $X=1$ when evaluated on the background \cite{Gubitosi:2012hu}. However, we will not consider such redefinition here in order to explicitly keep the kinetic term, $X$, in the final expressions for the EFT parameters.} \cite{Gubitosi:2012hu}
    \begin{align}
		c=&\,XK_{X},\\
		\Lambda=&\, XK_{X}-K,\\
		M_n^4=&\,(-X)^n\diffp[n]{K}{X},
    \end{align}
    for $n\geq2$. Therefore, the $k$-essence function $K$ contributes only to the mass parameters $M^4_n$ at perturbation level. So, there is no kinetic mixing between scalar and tensor perturbations.
	
	
    \subsection{Mass parameters for $G\Box\phi$\label{sec:appG}}
	
    From the definition of the d'Alembertian operator and the slicing vector $n_\mu$ (see equation (\ref{def:slicing})), it is straightforward to check that
    \begin{align}\label{eq:box}
		\Box\phi=-\epsilon\left(\sqrt{2X}\K+\frac{n^\mu}{\sqrt{2X}}\partial_\mu X\right).
    \end{align}
    To compute the EFT parameters for the braiding part of the KGB action (\ref{eq:actionKGB}), it will be useful to consider the following expansion in polynomials of $\delta X$  \cite{Gubitosi:2012hu}
    \begin{align}\label{eq:Gexp}
		G(\phi, \X)=\epsilon\sqrt{2\X}\sum_{m=0}^\infty l_m\left(\delta X\right)^m,
    \end{align}
    where $\X=X+\delta X$ is the perturbed kinetic term and
    \begin{align}\label{eq:Gexp_ln}
		l_m=\frac{1}{m!}\diffp[m]{}{\X}\left.\left(\frac{G(\phi,\X)}{\epsilon\sqrt{2\X}}\right)\right|_{\delta X=0},
    \end{align}
    are the coefficient in the expansion evaluated on the background. 
    Recall that we are not imposing a shift-symmetry in this section. Consequently, the above coefficients depend on the background values of both the scalar field and its kinetic term. That is $l_m=l_m(\phi,X)$.
   Let us emphasise again that this is an abuse of notation, since neither $\phi$ nor its kinetic term can explicitly appear in the unitary gauge. As mentioned in the previous section, background quantities as $\phi$ and $X$ are functions of time only, whereas scalar perturbations are encoded in time-time component of the metric. Hence, the perturbed braiding function, $G$, and the expansion coefficients, $l_m$,  should read $G(t,g^{00})$ and $l_m(t)$ in this gauge choice, respectively. Nevertheless, this abuse of notation will be convenient, where there is no risk of confusion, for obtaining the expressions of the EFT parameters in terms of $G(\phi,X)$ and its derivatives.

   Perturbing equation (\ref{eq:box}), and taking into account equations (\ref{eq:int lK}) and (\ref{eq:Gexp}), we obtain after few integrations by parts that
    \begin{align}
	&\int\textup{d}^4x\sqrt{-g}\,\left( -G\Box\phi\right)= \int\textup{d}^4x\sqrt{-g}\Bigg\lbrace-\left(\dot{X}l_0+2X\dot{l}_0\right)\sqrt{-g^{00}}\nonumber\\
		&+\sum_{n=1}^\infty\left(-X\right)^n\Bigg[ \left(2Xl_n+\frac{2n-1}{n} l_{n-1}\right)\K \nonumber\\
        &+ \left(\dot{X}l_n-\frac{\dot{l}_{n-1}}{n}\right)\sqrt{-g^{00}}\Bigg]\left(\delta g^{00}\right)^n\Bigg\rbrace,
    \end{align}
    up-to boundary terms; cf. with equation (86) in reference\textsuperscript{\ref{fn:Gub}} \cite{Gubitosi:2012hu}. The time derivative of the coefficients $l_n$ can be simplified by noting that
    \begin{align}
        \dot{l}_n=\dot{\phi}\diffp{l_n}{\phi}+(n+1)\dot{X}l_{n+1},
    \end{align}
    where it should be kept in mind that these coefficients are evaluated on the background.
    Since $\K=3H+\delta \K$, and expanding $\sqrt{-g^{00}}$ in powers of $\delta g^{00}$ as
    \begin{align}
      \sqrt{-g^{00}}=\sqrt{1-\delta g^{00}}=1-\sum_{n=1}^\infty \lambda_n(\delta g^{00})^n,
    \end{align}
    where
    \begin{align}\label{def:lambdaExpansion}
        \lambda_n\coloneqq\frac{(2n)!}{4^n(n!)^2(2n-1)},
    \end{align}
    we can readout the contribution of $-G\Box\phi$ to all terms in the effective action (\ref{actionEFT}). (Note that $G\Box\phi$ enters with a minus in the action (\ref{eq:actionKGB}), in contrast with the convention used in reference \cite{Gubitosi:2012hu}.) These are
    \begin{align}
        c=&\,XG_X\left(3H\dot\phi-\ddot{\phi}\right)-2XG_\phi,\\
        \Lambda=&\, XG_X\left(3H\dot\phi+\ddot{\phi}\right)  ,\\
        M_n^4=&\, n! \Bigg[a_n+3Hb_n+2\lambda_nX\left(G_\phi+G_X\ddot{\phi}\right)\nonumber\\
        &\left.-\sum_{m=1}^{n-1}\lambda_m a_{n-m}\right],\\
	\bar{m}_{n-1}^3=&-2b_{n-1},
    \end{align}
    for $n\geq2$, and where we have defined
    \begin{align}
		a_n\coloneqq&\,\frac{(-1)^{n+1}\dot{\phi}X^n}{n}\diffp{l_{n-1}}{\phi},\label{def:an}\\
		b_n\coloneqq&\,(-X)^n\left(2Xl_n+\frac{2n-1}{n}l_{n-1}\right),\label{def:bn}
    \end{align}
    with the coefficient $l_n$ from in equation (\ref{eq:Gexp_ln}). 
    Please note that the background parameters $c$ and $\Lambda$, and the first order mass parameters $M^4_2$ and $\bar{m}_1^3$ were already computed, for instance, in references \cite{Gubitosi:2012hu,Gleyzes:2013ooa} (mind the different conventions used there). (See also reference \cite{Cusin:2017mzw} for a discussion on the cubic and quartic order parameters in (\ref{actionEFT}).) Nevertheless, to the best of our knowledge, this is the first time these mass parameters are explicitly obtained in terms of the function $G$ and its derivatives for an arbitrary order. 
	

 \subsection{Mass parameters for the KGB theory\label{app:EFT FKGB}}

     The background quantities and mass parameters for the complete KGB theory (\ref{eq:actionKGB}) are given by the sum of the results presented in the two previous sections. These read
     \begin{align}
         c=& \, XK_X+XG_X\left(3H\dot\phi-\ddot{\phi}\right)-2XG_\phi \label{cEFT KGB} ,\\
        \Lambda=&\, XK_X-K+ XG_X\left(3H\dot\phi+\ddot{\phi}\right) \label{LambdaEFT KGB},\\
        M_n^4=&\, (-X)^n \diff[n]{K}{X} + n! \Bigg[a_n+3Hb_n+2\lambda_nX\left(G_\phi+G_X\ddot{\phi}\right)\nonumber\\
        &\left.-\sum_{m=1}^{n-1}\lambda_m a_{n-m}\right],\label{Mn4EFT KGB}\\
        \bar{m}_{n-1}^3=& -2b_{n-1}\label{m3n EFT KGB} ,
    \end{align}
    for $n\geq2$. As to be expected, the combinations $c+\Lambda$ and $c-\Lambda$ (see definitions (\ref{c_EFT}) and (\ref{Lambda_EFT})) coincide with the expressions for $\rho_\phi$ and $p_\phi$ for the non shift-symmetric KGB theory \cite{KGB1} (also compare with the shift-symmetric version introduced in equations (\ref{rho_phi}) and (\ref{p_phi}), respectively). In addition, the leading order ($n=2$) mass parameters read
    \begin{align}
        M_2^4=&\,X^2K_{XX}+\frac12 XG_X\left(3H\dot\phi+\ddot\phi\right)+3H\dot\phi X^2G_{XX}\nonumber\\
        &-X^2G_{\phi X}   ,\\
        \bar{m}_1^3=&\, 2\dot{\phi}XG_X.\label{eq:m13}
    \end{align}
    These results coincide with those presented in the literature modulo sign conventions and numerical factors in the definition of $X$; cf. with, for instance, references \cite{Gubitosi:2012hu,Gleyzes:2013ooa}. Moreover, in accordance with the results presented in Table 2 of reference \cite{Bellini:2014fua}, note that $M_2^4$ is connected  with the \textit{kineticity} term $\alpha_\textrm{K}$ defined in equation (\ref{eq:alphaK}) through $4M_2^4=H^2\alpha_\textrm{K}-2c$, whereas $\alpha_\textrm{B}$ is equal to $\bar{m}_1^3$ (modulo sign conventions); see definition in equation (\ref{eq:alphaB}).

    For the discussion in section \ref{sec:intGWs}, it is also important to compute the next-to-leading order braiding operator $\bar{m}_2^3$. This is
    \begin{align}
        \bar{m}_2^3=&- \frac12\dot{\phi}X\left(G_X+2XG_{XX}\right)\label{eq:m23}, 
    \end{align}
    as can be seen from equation (\ref{m3n EFT KGB}).  Comparing this expression with that of $\bar{m}_1^3$ in equation (\ref{eq:m13}), it follows that 
    \begin{align}
        4\bar{m}_2^3=-2\dot\phi XG_X\left(1+\frac{2XG_{XX}}{G_X}\right)=-\bar{m}_1^3\left(1+\frac{2XG_{XX}}{G_X}\right).
    \end{align}
    Consequently, the parameter $\eta$ used to measure deviations w.r.t. cubic Galileon (i.e. $G(X)\propto X$) in expression (\ref{etaEFT}) reads
    \begin{align}
        \eta=\frac{2XG_{XX}}{G_X},
    \end{align}    
    which is precisely the quantity first introduced in equation (\ref{def:eta}).
    
    Finally, for the shift-symmetric case of the KGB theory the functions $K$ and $G$ depend on the kinetic term $X$ but not on the scalar field itself. In this scenario, the expressions (\ref{cEFT KGB}) and (\ref{Mn4EFT KGB}) simplify as $G_\phi$ vanishes. In addition, the coefficients $a_n$ also become trivial since there is no dependence on $\phi$ in the shift-symmetric case; see definition (\ref{def:an}). Nevertheless, the formulas for the background quantity $\Lambda$ and the braiding-related parameters $\bar{m}_{n-1}^3$ remain absolutely the same.  Thus, the EFT parameters for the shift-symmetric KGB theory simply read
    \begin{align}
         c=&\,  XK_X+XG_X\left(3H\dot\phi-\ddot{\phi}\right)  ,\\
        \Lambda=&\, XK_X-K+XG_X\left(3H\dot\phi+\ddot{\phi}\right) ,\\
        M_n^4=&\, (-X)^n \diff[n]{K}{X} +n! \left(3Hb_n+2\lambda_n XG_X\ddot{\phi}\right),\\
        \bar{m}_{n-1}^3=& -2b_{n-1},
    \end{align}
   for $n\geq2$, where $\lambda_n$ and $b_n$ are defined in equations (\ref{def:lambdaExpansion}) and (\ref{def:bn}), respectively.

	
\bibliography{references}
	\bibliographystyle{elsarticle-num}

 
\end{document}